\documentclass[a4paper,11pt]{article}
\pdfoutput=1 
\usepackage{jcap/jcappub}

\usepackage{color}
\usepackage{graphicx}
\usepackage{amssymb}
\usepackage{hyperref}
\usepackage[utf8]{inputenc}

\newcommand{\dd}{\ensuremath{\textrm{d}}}

\newcommand{\const}{\textrm{const}}
\newcommand{\UD}[2]{\ensuremath{^{#1}_{\phantom{#1} #2}}}
\newcommand{\DU}[2]{\ensuremath{_{#1}^{\phantom{#1} #2}}}

\newcommand{\RR}{\ensuremath{{\mathbb{R}}}}

\newcommand{\beq}{\begin{equation}}
\newcommand{\eeq}{\end{equation}}
\newcommand{\bea}{\begin{eqnarray}}
\newcommand{\eea}{\end{eqnarray}}
\newcommand{\bit}{\begin{itemize}}
\newcommand{\eit}{\end{itemize}}
\newcommand{\bfi}{\begin{figure}}
\newcommand{\efi}{\end{figure}}
\newcommand{\bfic}{\begin{figure*}}
\newcommand{\efic}{\end{figure*}}
\newcommand{\bce}{\begin{center}}
\newcommand{\ece}{\end{center}}
\newcommand{\bt}{\begin{table}}
\newcommand{\et}{\end{table}}
\newcommand{\btb}{\begin{tabular}}
\newcommand{\etb}{\end{tabular}}

\newcommand{\du}{\ensuremath{\delta_u}}
\newcommand{\doo}{\ensuremath{\delta_{\mathcal{O}}}}

\newcommand{\calP}{\ensuremath{\mathcal{P}}}
\newcommand{\calD}{\ensuremath{\mathcal{D}}}
\newcommand{\calM}{\ensuremath{\mathcal{M}}}
\newcommand{\calN}{\ensuremath{\mathcal{N}}}
\newcommand{\calE}{\ensuremath{\mathcal{E}}}
\newcommand{\calO}{\ensuremath{\mathcal{O}}}
\newcommand{\calG}{\ensuremath{\mathcal{G}}}

\newtheorem{theorem}{Theorem}[section]

\newcommand{\qed}{\nobreak \ifvmode \relax \else
      \ifdim\lastskip<1.5em \hskip-\lastskip
      \hskip1.5em plus0em minus0.5em \fi \nobreak
      \vrule height0.75em width0.5em depth0.25em\fi}

\title{Optical drift effects in general relativity}

\author[a]{Miko\l{}aj Korzy\'nski}
\affiliation[a]{
Center for Theoretical Physics \\
Polish Academy of Sciences \\
Al. Lotnik\'o{}w 32/46, 02-668 Warsaw \\
Poland}
\emailAdd{korzynski@cft.edu.pl}

\author[b]{and Jaros\l{}aw Kopi\'nski}
\affiliation[b]{
Faculty of Physics \\
University of Warsaw \\
Pasteura 5, 02-093 Warsaw \\
Poland}
\emailAdd{jaroslaw.kopinski@fuw.edu.pl} 

\abstract{We consider the question of determining the optical drift effects in general relativity, i.e. the rate of change 
of the apparent position, redshift, Jacobi matrix, angular distance and luminosity distance of a distant object as registered by an observer in an arbitrary spacetime.
We present a fully relativistic and covariant approach, in which the problem is reduced to a hierarchy of ODE's solved along the line of sight.
The 4-velocities and 4-accelerations of the observer and the emitter and the geometry of the spacetime along the line of sight constitute the input data.
We build on the standard relativistic geometric optics formalism and extend it to include the time derivatives of the observables. 
In the process we obtain two general, non-perturbative relations: the first one between the gravitational lensing, represented by the Jacobi matrix, and the apparent position drift, also called the cosmic parallax, and the second one
between the apparent position drift and the redshift drift.
The applications of the results include the theoretical study of the drift effects of cosmological origin (so-called real-time cosmology) in numerical or exact Universe models.
}


\arxivnumber{1711.00584}

\begin{document}
\compress
\maketitle
\section{Introduction}

In the recent years we have witnessed the emergence of a new field in cosmology, often referred to as the ``real-time cosmology'', concerned with measuring small 
temporal changes of positions, 
redshifts and luminosity 
distances of objects at cosmological distances. Time variations of these quantities, also known as the optical drift effects, observed over
the time of $\approx 10$ years
can  provide important information
about the Universe on large scales and its evolution  \cite{quercellini}. This data would be independent from the standard cosmological 
observables like the CMB power spectrum, the current value of the Hubble parameter $H_0$ or the redshift-luminosity 
relation for supernovae, and thus could significantly expand our understanding of the Universe. 

Measuring the optical drift effects of cosmological origin is a challenging task for astronomers. The variations are most likely very small over the timescales of a single observational 
mission. The measurements therefore require instruments of enormous precision and a carefully crafted observational strategy combining data from many objects in order to increase the 
signal-to-noise ratio \cite{liske, Quercellini:2008ty, quercellini}. 
Nevertheless, some of the theoretical ideas have already been implemented: the Gaia satellite is measuring, among other things, the positions of $\approx 10^5$ distant quasars 
and may detect their variations over the mission time \cite{rasanen, quercellini, gaia1, quartin}. 
The European Extremely Large Telescope (E-ELT) will probe the spectra of distant quasars using an ultra-stable spectrometer CODEX, looking for variations in the Lyman-alpha forest
\cite{liske, quercellini, quartin, pasquini}. Similar measurements have been put forward with the ESPRESSO spectrograph and the VLT telescope \cite{Cristiani:2007by} as well as 
the Square Kilometer Array (SKA) \cite{Klockner:2015rqa}.

One of the reasons why researchers have been interested in the optical drift effects despite enormous observational difficulties  is that 
they potentially provide a direct, model-independent access to the history of the expansion of the Universe. Namely, in homogeneous FLRW models the values of the redshift drift 
$\dot z$ for objects at different redshifts $z$ are directly related to the Hubble parameter history $H(z)$ \cite{sandage, loeb, quercellini}.  
On the other hand, the position drift effects, or the ``cosmic parallax'', 
may probe large scale inhomogeneities of the Hubble flow or matter distribution \cite{quercellini,Quercellini:2008ty,Fontanini:2009qq,Quercellini:2009ni,Krasinski:2012ty}. 

From the theoretical point of view the drift effects are fairly easy to understand and calculate
in a perfectly homogeneous and isotropic FLRW Universe, however  in a less idealized model, containing inhomogeneities of different types, the drift 
effects are affected by the peculiar motions of the source (emitter) 
and of the observer, the local distortions of geodesics due to the variations of the local gravitational potential,
and, finally, by
the propagation effects through evolving inhomogeneities of various scales.

Similar problems arise when we consider the redshift-luminosity relation in realistic models of the Universe: the light propagation through the inhomogeneous
structure differs from the propagation through a perfectly homogeneous spacetime 
~\cite{Fleury:2014gha,
1967ApJ150737G,
1969ApJ...155...89K,
1970ApJ...159..357R,
Lamburt2005,
2010PhRvD..82j3510B,
2011PhRvD..84d4011S,
Nwankwo:2010mx,
2012MNRAS.426.1121C,
2012JCAP...05..003B,
Lavinto:2013exa,
Troxel:2013kua,
Bagheri:2014gwa,
Peel:2014qaa}
and realistic models of the Hubble diagrams must take into account
the inhomogeneity of matter distribution on fine scales, especially the influence of the reduced Ricci focusing in the empty regions \cite{zeldovich, dyer-roeder, Dyer:1973zz}. 
From the mathematical point of view the approximations and models discussed in most of the above mentioned papers are based on 
the geometric optics and gravitational lensing formalism in general relativity, introduced by Sachs \cite{sachs} and developed later by many authors 
\cite{ehlers-jordan-sachs, seitzschneiderehlers}. The formalism uses the geodesic deviation equation (GDE), either directly or in its equivalent formulation as  
the Sachs optical equations, to extract  the information about the gravitational lensing as well as the angular and luminosity distances 
from an observer to distant objects. 
See \cite{perlick-lrr} for a longer review and references to the most important papers in the field.

Unfortunately, there seems to be no comparable formalism designed for the optical drifts effects in general relativity (GR). 
Some of the theoretical work on the position drift effects has been based on concrete classes of Universe models 
   \cite{novikov-popova, Krasinski:2011iw, Krasinski:2012nw}, including the Lema\^itre-Tolman-Bondi (LTB) or Szekeres solutions \cite{ Quercellini:2008ty, Krasinski:2010rc, Krasinski:2012ty}. 
Other papers use 
rather simplified geometric arguments to obtain the results \cite{Quercellini:2008ty,quercellini}. Finally some authors  
proposed  a more general and  mathematically refined approach to the problem
\cite{mccrea,weinberg-letter,rosquist,kasai,rasanen}, but in the expressions they provide
the dependence of the result on the observer's and emitter's motion is not explicit. Most  authors  assume  the existence of a well-defined large-scale cosmological flow,
including sometimes the possibility of non-relativistic peculiar motions. 
Piattella and Giani  noted recently  \cite{Piattella:2017uat} the relation between gravitational lensing (considered in thin lens approximation) and 
the position drift as well as the signal arrival time drift in  FLRW models with a pointlike lens, but to our knowledge no work so far considered  the general, quantitative relation between the gravitational 
lensing and the drift effects. 

The recent paper of Hellaby and Walters \cite{Hellaby:2017soj} offers the most general approach
so far and provides expressions for the redshift, angular distance, density of objects in the redshift space and an
expression for the position drift (called the ``proper motion'') as registered by an arbitrary observer and in any metric. The authors use the Fermi-Walker transport
of 
the observer's orthonormal basis to define fixed directions in the sky (just like in this paper) and express their results using a specific coordinate system, i.e. the past null cone coordinates, 
as well as the various transport equations 
for the basis along null geodesics. They  do not assume any  relation of the observer and emitter to the Hubble flow, but they also do not discuss explicitly the dependence
on the observer's 4-velocity or 4-acceleration (the dependence is in fact present in the initial conditions for the ODE's the authors solve). 
Nevertheless the formalism presented in \cite{Hellaby:2017soj} seems to be very useful in numerical 
investigation of concrete inhomogeneous models and the authors apply it to the Szekeres solution. 

In the literature  concerning the redshift drift the situation is similar: some papers consider the effects in known exact solutions
\cite{Yoo:2008su, Quartin:2009xr, Yoo:2010hi} or specialized classes \cite{Uzan:2008qp}. Gasperini, Marozzi, Nugier and Veneziano derived the redshift
drift in a general spacetime for a geodesic observer, expressed in the past null cone coordinates \cite{Gasperini:2011us}. 
None of these papers discusses the general dependence of the effect on the peculiar motions of the emitter and observer or 
its relation to other drift effects. Moreover, we are not aware of any  theoretical  papers  about the luminosity or angular distance drift in a general situation.

In this paper we would like to fill these gaps and present
an exact analysis of the most general situation, in which two bodies, an emitter and an observer, are placed in an arbitrary spacetime. The light propagation is treated within the
geometric optics approximation and the light rays in a neighbourhood of a known null geodesic are treated using the GDE.
In order to keep the formalism as general as possible, we assume no splitting  of the spacetime geometry or the matter flow into the ``background'' and ``perturbations''. 
Likewise, we do not refer to any large-scale
Hubble flow of matter, but rather consider the dependence of the result on the momentary motions of the emitter and the observer, both of which we consider arbitrary. 
We build a hierarchy of ordinary differential equations (ODE's) one needs to solve along a null geodesic in order to estimate the drift effects registered
by the observer on the geodesic's future end. 
The hierarchy begins with the standard null GDE, needed to evaluate Jacobi matrix related to the lensing effects of the spacetime.
It is followed by an inhomogeneous ODE required to obtain the position drift at the observer's sky. Next we may use the data from the previous steps to obtain the redshift drift.
Finally, we need to solve another set of inhomogeneous ODE's, again using the data from the previous steps, to obtain the Jacobi matrix drift and the drifts of the luminosity and 
angular distances. Thus, the problem of evaluating the optical observables and their drifts in a spacetime is reduced to solving ODE's. The ODE's take the 4-velocities and 
the 4-accelerations of the emitter and observer as well as the Riemann tensor and its first derivative as their input data.
All equations are formulated in a covariant,
geometric way.

The advantage of this approach is threefold: firstly, equations formed as a hierarchy are easier to implement numerically, since we may use the data from the previous steps
in the next steps. Secondly, the results
we derive relate various observable quantities
and their drifts to each other. They can therefore be seen as non-perturbative identities in geometric optics, akin to the celebrated reciprocity relations discovered by Etherington 
\cite{etherington, etherington2}. Thirdly, in the formulas we derived for the position and redshift drift we managed to separate cleanly 
the dependence on the observer's and emitter's motion and
the dependence on the spacetime geometry.

Possible applications of the results presented in this paper include numerical relativity: the formulas for the drift effects can be used together with the standard Sachs equations
to calculate the drift registered by any observer of any source by solving ODE's along a null geodesic, extending the existing numerical studies of 
the optical properties of inhomogeneous cosmological 
models \cite{Bentivegna:2016fls, Sanghai:2017yyn}.
Moreover, the same formulas provide the starting point for studying the drift effects from inhomogeneities using stochastic methods, in similar manner to
the approach used by Fleury, Larena and Uzan to the redshift-luminosity relations \cite{Fleury:2015rwa}. Finally, it is worth investigating whether the formalism
may be used to prove results
concerning exact cosmological models, including potentially 
the characterization of the FLRW metric using the optical drift effects, in analogy with the well-known theorem by Ehlers, Geren and Sachs and its generalizations 
\cite{ehlers-geren-sachs,stoeger}. The formalism is of course valid beyond the cosmological context and may find applications in the study of time-dependent 
gravitational lensing on smaller distances.

\paragraph{Limits of applicability. } The formalism presented in this paper is fully covariant and valid for arbitrary spacetime geometry. In particular, the equations derived 
here work for both strong and weak lensing and in both weak and strong gravitational fields. In the context of cosmology they are applicable 
to both linear and nonlinear inhomogeneities. The approach is also valid for arbitrary  motions of the sources and the observer. In particular it works for
relativistic or ultrarelativistic motions with respect
to the local cosmological flow.
The only limitations of the formalism are due to the limits of applicability of the
geometric optics approximation, which in the astronomical and cosmological context is known to work very well, and  of the geodesic deviation equation. The latter
is valid under the assumption that the curvature scale and the size of focusing or defocussing inhomogeneities are much larger than the width of the light beam considered. In this paper we also need to assume that the emitter does not lie on a caustic formed by the observer's past light cone.

\paragraph{Article structure. }The article is organized as follows: we begin the Section 2 by revisiting the geometric optics in GR and discussing the classical results. 
In our approach we emphasize the covariance and  the observer-independence of the formalism. 
In the meantime we introduce the necessary mathematical machinery and  the notation needed in the rest of the paper. 
The main new results of the paper are stated in Section 3, in which we derive the general formulas the 
position, redshift, Jacobi matrix and angular as well as luminosity distance drifts. In Section 4 we discuss the dependence of the drift quantities on the  
observer's and emitter's motion, relate the position drift expression to the cosmic parallax as discussed in the literature and  conclude with a summary of the results.

\section{Geometric optics in GR}  \label{sec:geometricoptics}

We begin this section by recalling the standard results concerning the geometric optics in general relativity 
formulated in a geometric, coordinate-independent manner. The main mathematical tool used in this context is the null geodesic deviation equation, 
which describes the change of shape of an infinitesimally thin beam 
of null geodesics. The methods and results discussed here, developed  by 
Sachs \cite{sachs}, Jordan, Ehlers and Sachs \cite{ehlers-jordan-sachs} (reprinted in \cite{ Jordan2013reprint}),  are very well-known, but we present them here 
in order to make the paper self-contained and 
because the new results should be understood as a natural extension of the standard formalism. In our exposition we will emphasize the observer-independence 
and covariance of the formalism and  
discuss how one may distinguish between the observer- and emitter-independent effects of wave propagation through curved spacetime and the effects
due to the peculiar motions. Interested readers may also see \cite{perlick-lrr, seitzschneiderehlers} for a longer review of the topic.

\subsection{Geodesics and beams} 

%comment - zmieniam \gamma na \gamma_0, bo w nastepnym zdaniu jest gamma_0
% reply - OK
Let $M$ be the spacetime with a metric $g$ of signature $(-,+,+,+)$. Let $\gamma_0$ be an affinely parametrized geodesic given by $x^\mu(\lambda)$.
Consider also a one-parameter family of geodesics $\gamma_\varepsilon$ such that $\varepsilon =0$ corresponds to the original, fiducial geodesic. 
This defines a mapping $F: (\varepsilon,\lambda) \to \gamma_\varepsilon(\lambda)$
from an open subset of $\RR^2$ to $M$. In the generic case the image of the mapping constitutes at least locally a 2-surface in $M$. 

\bfi
%\bce
\centering
%\begin{minipage}[b]{0.65\linewidth}
\includegraphics[width=0.5\textwidth]{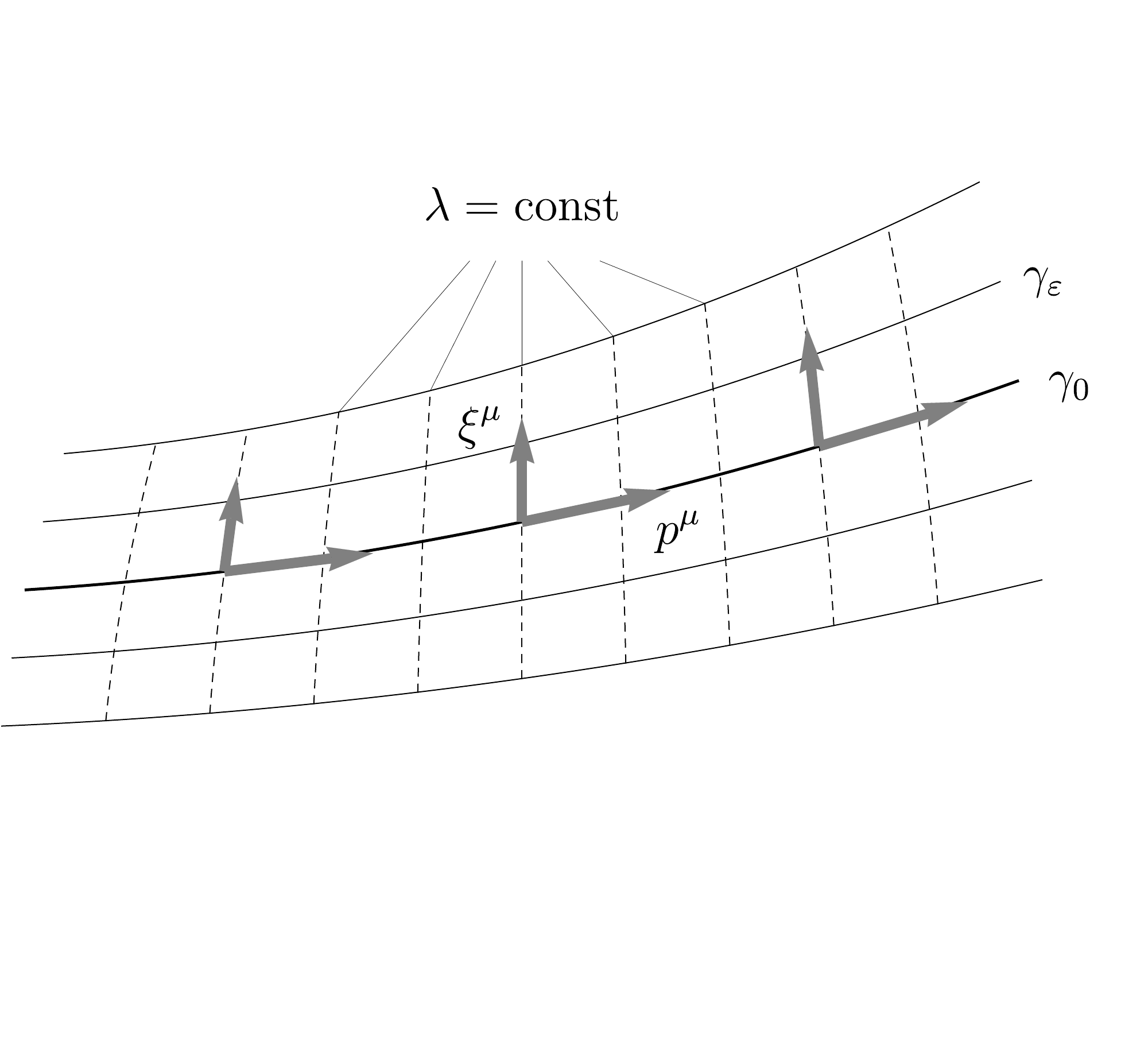}
%\end{minipage}
\caption{Vectors $\xi^\mu$ and $p^\mu$ defined by the one-parameter family of geodesics $\gamma_\varepsilon$.}
%An observer with four-velocity $u^\mu$ along the null geodesic $\gamma_0$, the backward-pointing 4-momentum of the photon $p^\mu$,
%the 3 remaining Sachs frame vectors: $r^\mu$, pointing in the direction from which the photon is coming and the screen vectors $e\DU{A}{\mu}$. %The spatial
%3-space $u_\perp$ denotes the plane of simultaneity of the observer.

\label{fig:lambdaepsilon}
%\ece
\efi
%comment - zmienilem geodesic gamma_e na family gamma_e, bo moglo byc mylace
%reply - OK
The pushforward of the vector $\partial_\lambda$ is the tangent vector field to the family $\gamma_\varepsilon$. On the
other hand, the pushforward $\xi = F_*(\partial_\varepsilon)$ is usually called the deviation vector. It connects the points of the same value of $\lambda$ on infinitesimally 
close geodesics, see Fig. \ref{fig:lambdaepsilon}. More precisely, in the first order of expansion in $\varepsilon$ around $\gamma_0$ we have
\bea
x^\mu(\varepsilon,\lambda) = x^\mu(\lambda)\Big|_{\varepsilon=0} + \varepsilon\,\xi^\mu(\lambda) + O(\varepsilon^2) \label{eq:varepsilonexpansion}
\eea
in any coordinate system $(x^\mu)$. The vectors $\partial_\lambda$ and $\partial_\varepsilon$ commute, therefore so do their pushforwards:
\bea
\nabla_\xi p^\mu - \nabla_p \xi^\mu = 0. \label{eq:commutation}
\eea
The l.h.s. of the equation above involves only the covariant derivatives along the 2-surface made of the  geodesics, therefore 
it is well defined even though the vector fields $p^\mu$ and $\xi^\mu$ are undefined outside the 2-surface.

\subsection{Geodesic deviation equation}
%comment - czy pierwsze dwa zdania sa tutaj potrzebne? Od poczatku wystartowalismy z zalozenia o tym, ze krzywe sa geodezyjnymi
%reply - zostawić
Equations (\ref{eq:varepsilonexpansion}--\ref{eq:commutation}) are satisfied for any surface made of sufficiently regular curves, not necessary geodesics. 
We have assumed $\gamma_\varepsilon$
to be geodesics, i.e. to satisfy additionally
\bea
\nabla_p p^\mu = 0 \label{eq:geod}
\eea
for all $\varepsilon$. We differentiate this equation with respect to $\xi^\mu$ to obtain the condition for $\xi^\mu$ under which (\ref{eq:geod}) remains true:
\bea
0 = \nabla_\xi\left(\nabla_p p^\mu\right)& =& \nabla_p\left(\nabla_\xi p^\mu\right) + R\UD{\mu}{\kappa\sigma\rho}\,p^\kappa\,\xi^\sigma\,p^\rho + \left(\left[\xi,p\right]\right)^\sigma
\,\nabla_\sigma p^\mu \nonumber
\eea
(we have used the definition of the Riemann curvature tensor here). Now we note that because of (\ref{eq:commutation}) the last term containing the Lie bracket vanishes and we
are left with
\bea
\nabla_p\left(\nabla_\xi p^\mu\right) + R\UD{\mu}{\kappa\sigma\rho}\,p^\kappa\,\xi^\sigma\,p^\rho = 0.  \nonumber
\eea
Making use of (\ref{eq:commutation}) again and shuffling the indices, we obtain the  geodesic deviation equation (GDE), also known as the Jacobi equation:
\bea
\nabla_p\nabla_p \xi^\mu - R\UD{\mu}{\alpha\beta\nu}\,p^\alpha\,p^\beta\,\xi^\nu &=& 0 \label{eq:GDEfirst}.
\eea
This is a second order ODE for $\xi^\mu$ along $\gamma_0$. In order to simplify the notation, we introduce the geodesic deviation differential operator
\bea
\calG[\xi]^\mu = \nabla_p\nabla_p \xi^\mu - R\UD{\mu}{\alpha\beta\nu}\,p^\alpha\,p^\beta\,\xi^\nu \label{eq:GDopdef}.
\eea
Now (\ref{eq:GDEfirst}) can be given a simpler form:
\bea
\calG[\xi]^\mu = 0.
\eea

Equation (\ref{eq:GDEfirst}) has been derived under the assumption that we are given a one-parameter family of curves all of which are geodesic.
However, it can also be interpreted differently. Assume that we are given  only
a single geodesic  $\gamma_0$ and that the GDE (\ref{eq:GDEfirst}) holds along $\gamma_0$ for a given vector field $\xi^\mu$. 
We can then consider a one-parameter family of curves $\gamma_\varepsilon$, given by (\ref{eq:varepsilonexpansion}), for small $\varepsilon$. In this case we can check that for all 
curves $\gamma_\varepsilon$ we have
\bea
\nabla_p p^\mu = O(\varepsilon^2),  \nonumber
\eea
i.e. curves $\gamma_\varepsilon$ remain geodesic up to the linear order in $\varepsilon$. Thus, the GDE provides the 
first order condition for a variated geodesic curve to remain geodesic.

\subsection{Properties of the GDE}
Now we will note two important properties of the GDE. These properties hold irrespectively of the underlying geometry, i.e. independently of the 
form of the components of the Riemann tensor. First we multiple (\ref{eq:GDEfirst}) by $p_\mu$ and obtain
\bea
\nabla_p \nabla_p (p_\mu\,\xi^\mu) = 0  \nonumber
\eea
which can be solved immediately:
\bea
p_\mu\,\xi^\mu = A + B\,\lambda. \label{eq:ABlineardef}
\eea
This way we have defined two conserved quantities of the GDE: for any solution $\xi^\mu(\lambda)$
\bea
B &=& (\nabla_p\xi^\mu)\,p_\mu = \const, \label{eq:Bconst} \\
A &=& \xi^\mu\,p_\mu - B\,\lambda = \const. \label{eq:Aconst}
\eea
On the other hand, it is also easy to see that any transformation of the form
\bea
\xi^\mu(\lambda) \to \bar\xi^\mu(\lambda) = \xi^\mu(\lambda) + (C + D\,\lambda)\,p^\mu \label{eq:gauge}
\eea
with $C,D=\const$ transforms solutions into solutions. Transformations of this kind have a simple geometric interpretation: they correspond to the affine reparametrizations
of the neighbouring geodesics
around $\gamma_0$, without affecting $\gamma_0$ itself (see Fig. \ref{fig:gauge}). Geometrically, $\bar\xi^\mu(\lambda)$ corresponds to the same congruence of geodesics as $\xi^\mu$, 
but with a change 
of parametrization around $\gamma_0$ not affecting
the fiducial geodesic itself.
\bfi
%\bce
\centering
%\begin{minipage}[b]{0.65\linewidth}
\includegraphics[width=0.5\textwidth]{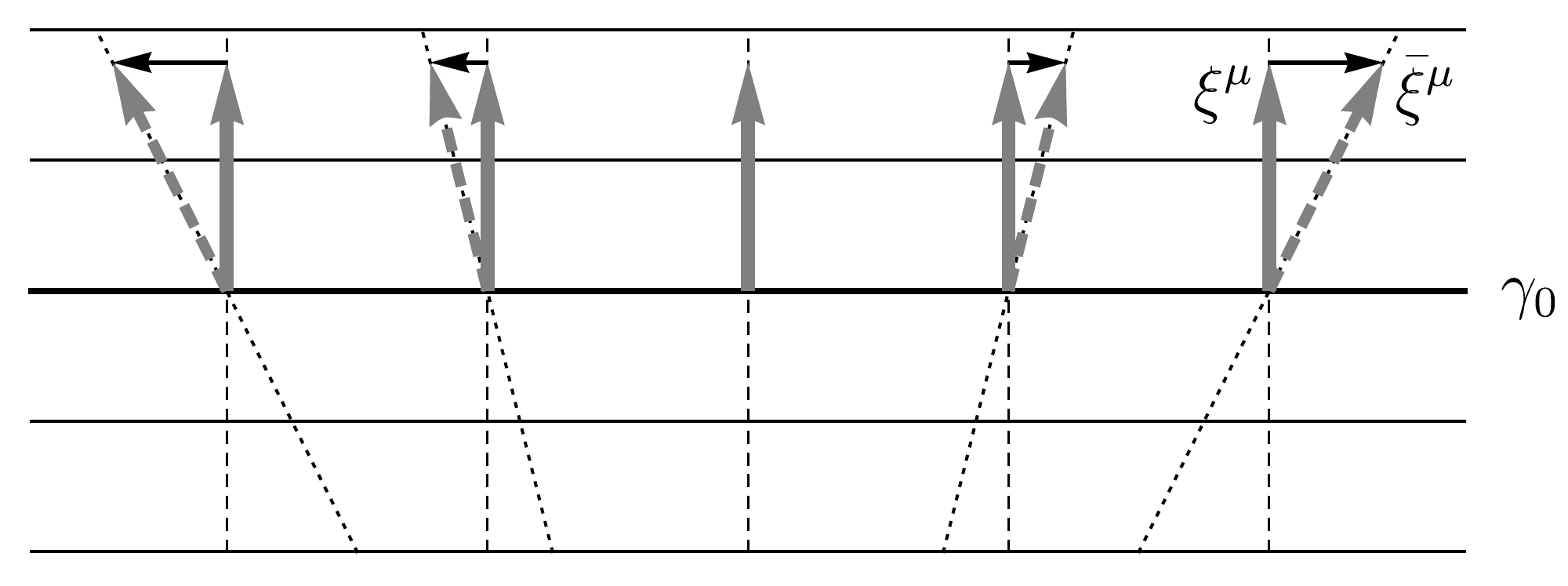}
%\end{minipage}
\caption{Gauge transform (\ref{eq:gauge}) as an affine reparametrization in the infinitesimal neighbourhood of $\gamma_0$.}
%An observer with 4-velocity $u^\mu$ along the null geodesic $\gamma_0$, the backward-pointing 4-momentum of the photon $p^\mu$,
%the 3 remaining Sachs frame vectors: $r^\mu$, pointing in the direction from which the photon is coming and the screen vectors $e\DU{A}{\mu}$. %The spatial
%3-space $u_\perp$ denotes the plane of simultaneity of the observer.

\label{fig:gauge}
%\ece
\efi
The observations of this subsection obviously shows that while GDE,  as a second order ODE for a vector of dimension 4, has an 8-dimensional space of solutions,
the true, physical dimension of the space of solution is lower.

\subsection{Null GDE}\label{sec:nullgde}
Consider GDE around a null geodesic $\gamma_0$, i.e. such that 
\bea
p^\mu \,p_\mu = 0. \label{eq:null}
\eea
Assume that we only consider a beam of geodesics which are all null, i.e. the perturbed geodesics also satisfy (\ref{eq:null}).
% comment
%całe następne zdanie w nawiasie?
%relpy - skasowałem nawias
% 
In the language of differential geometry we are considering  a congruence of null geodesics. This means that the deviation vector must satisfy the condition
\bea
(\nabla_\xi p^\mu)\,p_\mu = 0,  \nonumber
\eea
or equivalently, if we use (\ref{eq:commutation}), 
\bea
(\nabla_p \xi^\mu)\,p_\mu = 0\label{eq:null2}.
\eea
Note that it is enough if we impose (\ref{eq:null2}) at a single point along $\gamma_0$: the GDE implies (\ref{eq:Bconst}) and the condition propagates to every point. 
It is also easy to see from (\ref{eq:Aconst}) that in this case $A = \xi^\mu\,p_\mu = \const$, i.e. the product of $p^\mu$ and $\xi^\mu$ is conserved. In fact, the 
converse is also true: assume that the value of $\xi^\mu\,p_\mu$ at a point $\lambda_1$ is equal to its value at any different point $\lambda_2$. 
Then from (\ref{eq:ABlineardef}) we see that $B = 0$ and $A=\const$, which implies
(\ref{eq:null2}). We will make use of this fact later.

We will now focus on a special situation when the perturbed null geodesic, remaining null, satisfies additionally the condition $A=0$, or
\bea
  \xi^\mu\,p_\mu = 0, \label{eq:xiortp}
\eea
i.e. the deviation vector is perpendicular to $p^\mu$. We immediately note two things:
\begin{enumerate}
\item Condition (\ref{eq:xiortp}) is insensitive to adding to $\xi^\mu$ anything proportional $p^\mu$. In particular, 
it is gauge-invariant in the sense of transformations (\ref{eq:gauge}), i.e. it is a property of the perturbed geodesic insensitive
to its parametrization,
\item Note that again if (\ref{eq:xiortp}) is assumed to hold at one point along
$\gamma_0$, it is satisfied everywhere thanks to (\ref{eq:null2}). 
\end{enumerate}

In order to elucidate the geometric meaning of condition (\ref{eq:xiortp}), we will fix an observer $u^\mu$ at a point along
$\gamma_0$. The observer introduces his \emph{adapted frame}, i.e. an orthonormal frame $\left(u^\mu,e\DU{A}{\mu},r^\mu\right)$ in which the timelike vector coincides with $u^\mu$ and
the two spatial vectors 
$e\DU{A}{\mu}$, $A=1,2$, are orthogonal to $p^\mu$. 
The third spatial vector $r^\mu$ points in the direction from
which the photon comes in the observer's frame, i.e.
\bea
p^\mu = Q(-u^\mu + r^\mu) \label{eq:sachsrE}
\eea
%comment - dluzszy komentarz o tym, czym są Q i a?
%reply - nie ma po co, te wielkości nie mają oczywistego sensu fizycznego
with a positive proportionality constant $Q > 0$, see Fig.\ref{fig:Sachsframe}. $Q$ is proportional to the energy of the photon as measured by the observer $u^\mu$
\bea
E_u = a\,Q = a\,p_\mu\,u^\mu, \label{eq:sachsenergy}
\eea
with the positive proportionality constant $a$ depending of the affine parametrization of $\gamma_0$, but the same
for all possible observers along $\gamma_0$.  
The vectors $e\DU{A}{\mu}$, called the \emph{Sachs basis}, span the 2-dimensional spatial subspace orthogonal to the direction of propagation of the photons, referred to as the
\emph{screen space}. Note that for a given observer $u^\mu$ and a given null vector $p^\mu$ the only remaining freedom of choosing the adapted frame is the 
freedom to rotate the Sachs basis around $r^\mu$.

\bfi[!h]
%\bce
\centering
%\begin{minipage}[b]{0.65\linewidth}
\includegraphics[width=0.5\textwidth]{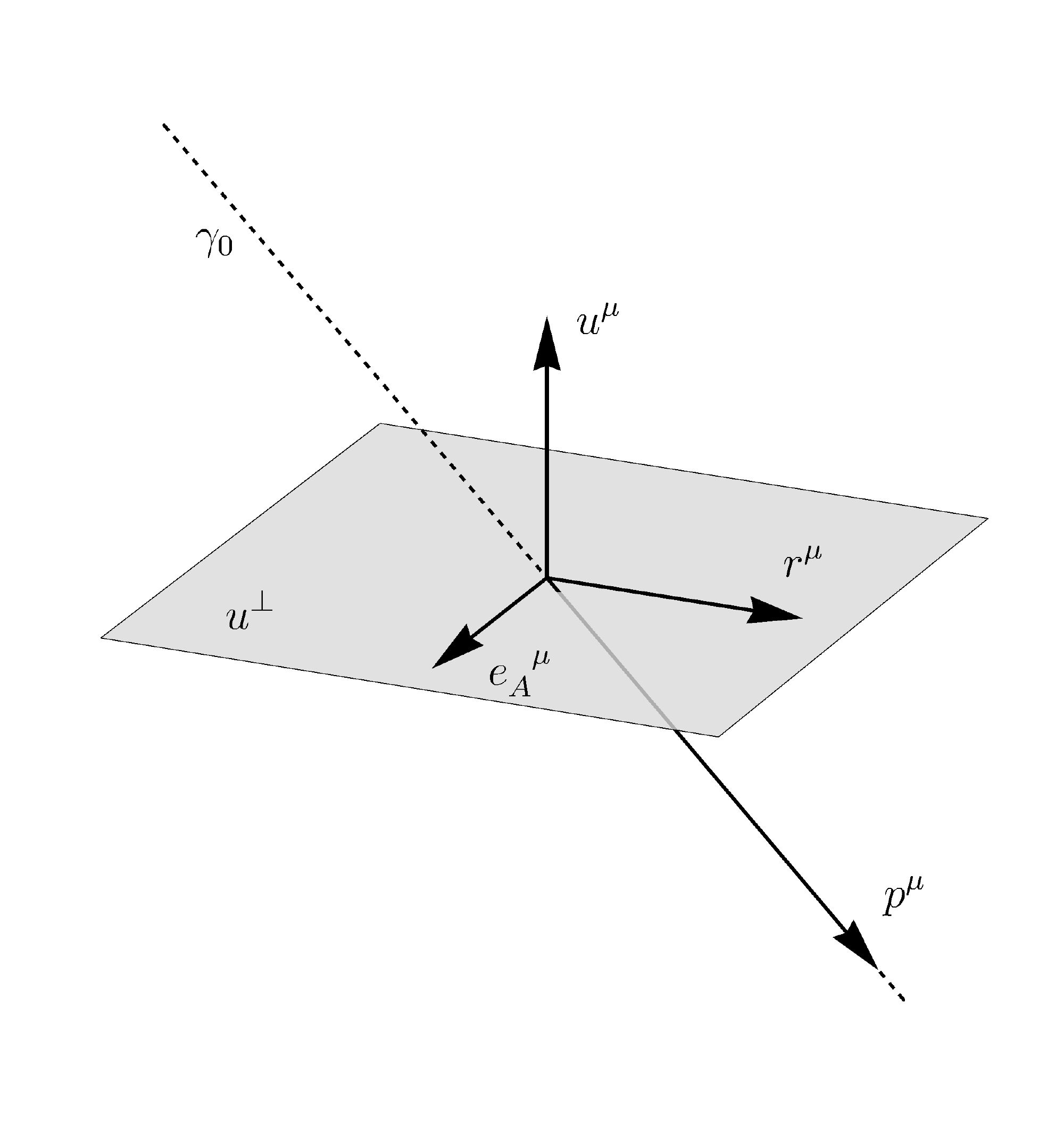}
%\end{minipage}
\caption{An observer with 4-velocity $u^\mu$ along the null geodesic $\gamma_0$, the backward-pointing 4-momentum of the photon $p^\mu$, the 3 remaining Sachs frame 
vectors: $r^\mu$, pointing in the direction from which the photon is coming and the screen vectors. The spatial
3-space $u^\perp$ denotes the plane of simultaneity of the observer.}
%An observer with 4-velocity $u^\mu$ along the null geodesic $\gamma_0$, the backward-pointing 4-momentum of the photon $p^\mu$,
%the 3 remaining Sachs frame vectors: $r^\mu$, pointing in the direction from which the photon is coming and the screen vectors $e\DU{A}{\mu}$. %The spatial
%3-space $u_\perp$ denotes the plane of simultaneity of the observer.

\label{fig:Sachsframe}
%\ece
\efi

Consider a photon travelling along a nearby geodesic $\gamma_\epsilon$ with $\epsilon \ll 1$. Since the observer is located exactly at a point on 
$\gamma_0$, the vectors $x_\epsilon^\mu = \epsilon(\xi^\mu + C\,p^\mu)$ 
 represent the positions of the nearby photon $\gamma_\epsilon$ with respect to the observer at various moments as it passes nearby, 
see (\ref{eq:varepsilonexpansion}).
Out of these vectors we single out the one that corresponds to the position at the moment simultaneous with the passing of 
the $\gamma_0$ photon through the observer as reported in the observer's frame (see Fig. \ref{fig:Sachsprojection}), denoted $\tilde x_\epsilon^\mu$. It needs to satisfy
\bea
\tilde x_{\epsilon}^\mu\,u_\mu = 0.  \nonumber
\eea
The only solution is
\bea
\tilde x_{\epsilon}^\mu = \epsilon\,\tilde\xi^\mu,   \nonumber
\eea
where $\tilde\xi^\mu$ is the projection of any $\xi^\mu$ to $u_\perp$ along the direction of $p^\mu$:
\bea
\tilde \xi^\mu &=& \xi^\mu + D\,p^\mu  \nonumber\\
\tilde \xi^\mu \,u_\mu &=& 0.  \nonumber
\eea

If we assume that the original $\xi^\mu$ is perpendicular to $p^\mu$, then so is $\tilde\xi^\mu$.  
This in turn is equivalent to the condition that $\tilde\xi^\mu$ is perpendicular to the direction of propagation $r^\mu$. Therefore, 
the observer sees the nearby photon located on his screen space spanned by $e\DU{A}{\mu}$ (see Fig. \ref{fig:wavefront}). Note that this statement is \emph{independent of the choice of the observer $u^\mu$}:
we have assumed nothing special about him, so the reasoning should hold for every observer along $\gamma_0$ and with any 4-velocity
as long as (\ref{eq:xiortp}) is true; every observer will agree that the new photon lies on his screen plane at the moment $\gamma_0$ passes him. 
We will thus say that \emph{$\xi^\mu$ satisfying (\ref{eq:xiortp}) corresponds to a photon displaced orthogonally with respect to $\gamma_0$\footnote{From the point of view of the wave optics $\xi^\mu$ is tangent to the wavefront passing through a given point, i.e. the set of points
with the same wave phase \cite{perlick-lrr}.}.
} The subspace of vectors perpendicular to $p^\mu$ at a given point, containing all the 
photons displaced orthogonally, shall be denoted by $p^\perp$.
\bfi
%\bce
\centering
%\begin{minipage}[b]{0.65\linewidth}
\includegraphics[width=0.5\textwidth]{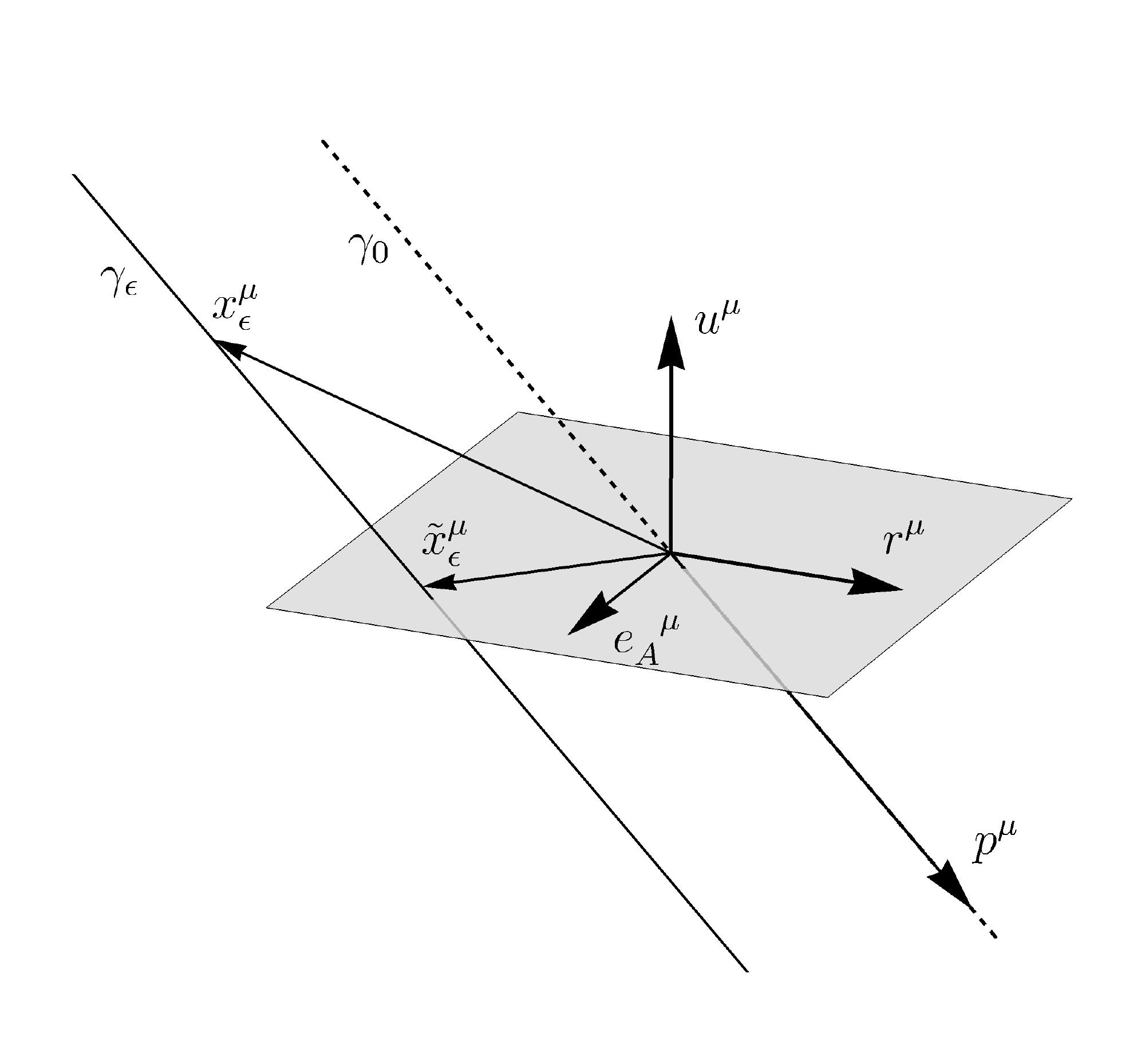}
%\end{minipage}
\caption{The position vector $\tilde x_\epsilon^\mu$ of the other photon $\gamma_\epsilon$ registered by the observer in the moment he/she is passed by the fiducial photon $\gamma_0$.}
%An observer with 4-velocity $u^\mu$ along the null geodesic $\gamma_0$, the backward-pointing 4-momentum of the photon $p^\mu$,
%the 3 remaining Sachs frame vectors: $r^\mu$, pointing in the direction from which the photon is coming and the screen vectors $e\DU{A}{\mu}$. %The spatial
%3-space $u_\perp$ denotes the plane of simultaneity of the observer.
\label{fig:Sachsprojection}
%\ece
\efi

\bfi
%\bce
\centering
%\begin{minipage}[b]{0.65\linewidth}
\includegraphics[width=0.5\textwidth]{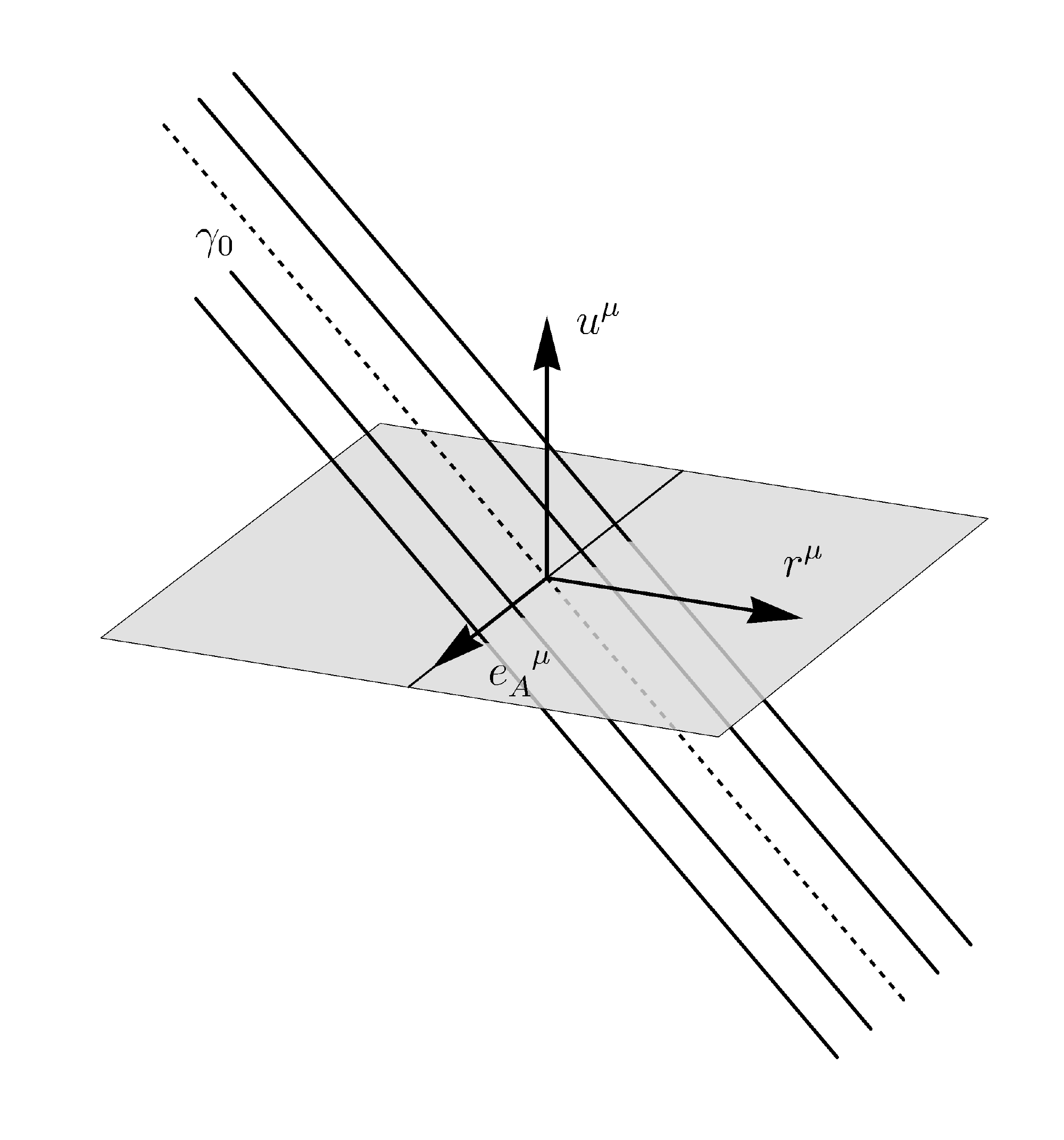}
%\end{minipage}
\caption{Photons displaced orthogonally with respect to $\gamma_0$.}
\label{fig:wavefront}
%\ece
\efi

Even with condition (\ref{eq:xiortp}) imposed we still have the freedom of adding $p^\mu$ multiplied by an affine function of $\lambda$ to solutions $\xi^\mu$,
as noted in equation (\ref{eq:gauge}). The reason is that $p^\mu$ is null and therefore self-orthogonal. Physically, this corresponds to a gauge transformation, because
the geodesic paths remain unaffected. We can get rid of this superfluous degree of freedom by considering the quotient space $\calP = p^\perp / p$, 
consisting of vectors in $p^\perp$ divided by the relation $x^\mu \sim y^\mu \iff x^\mu = y^\mu + C\,p^\mu$. The resulting space is two-dimensional. 

\paragraph{Notation. } The vector in $\calP$ corresponding to the equivalence class of $x^\mu \in p^\perp$ will be denoted by a boldface letter $\boldsymbol{x}$, and its components
will be denoted by a capital Latin index $A$ running from 1 to 2.

In the rest of this subsection we will argue that $\calP$ is also the most natural setting to consider the 
null GDE and geometric optics in general relativity.
Firstly, we note that the scalar product of two vectors $a^\mu$ and $b^\mu$ perpendicular to $p^\mu$ is insensitive to adding $p^\mu$ to any of them, i.e.
\bea
a^\mu\,b^\nu\,g_{\mu\nu} = (a^\mu + E\,p^\mu)\,(b^\nu + F\,p^\nu)\,g_{\mu\nu}.  \nonumber
\eea
This observation means that we can pull back the metric tensor from $p^\perp$ to $\calP$ and consider simply the
product of the equivalence classes $\boldsymbol{a}^A$, $\boldsymbol{b}^A$:
\bea
a^\mu\,b^\nu\,g_{\mu\nu} = \boldsymbol{a}^A\,\boldsymbol{b}^B\,\boldsymbol{g}_{AB}.  \nonumber
\eea
The tensor $\boldsymbol{g}_{AB}$ is positive-definite and thus $\calP$ is endowed in a natural way in a positive scalar product.

Secondly, we note that while passing from the full solution $x^\mu$ of the GDE to $\boldsymbol{x}^A$ involves loss of information concerning the dependence of the position of the photon 
on the affine parameter along its trajectory, this is irrelevant when we consider only the geometric optics. This can be seen as follows.

Let $u^\mu$ correspond again to an observer
at a point along $\gamma_0$ at $\lambda=\lambda_u$ and let $\tilde x_\epsilon^\mu$ be  the position of the other photon in 
the moment of his or her simultaneity, i.e. $\tilde x_\epsilon^\mu  =  x_\epsilon^\mu(\lambda_u) + C\,p^\mu$ additionally satisfying
$\tilde x_\epsilon^\mu\,u_\mu = 0$. We know that the position of the $x^\mu$ photon is perpendicular to the direction of propagation, so we know that
in the observer's Sachs frame it must be a combination of his screen vectors: 
$\tilde x_\epsilon^\mu = \tilde x^A\,e\DU{A}{\mu}$. The two components $\tilde x^A$, containing all information about the geometry of the 
light rays as recorded by $u^\mu$, can be evaluated as
\bea
\tilde x_A = e\DU{A}{\mu} \,(\tilde x_\epsilon)_\mu  \nonumber
\eea
(the quantities with lower and upper index $A$ being equal in positive signature).
Note that in order to calculate the components $\tilde x_A$ in the adapted frame, we only need the \emph{equivalence class} of $\tilde x_\epsilon^\mu$, i.e. $\tilde{ \boldsymbol{x}}_\epsilon$, and the equivalence classes
of the screen vectors. It is easy to see that the equivalence class $\tilde {\boldsymbol{x}}_\epsilon$  is equal to the 
equivalence class of the original $x_\epsilon^\mu$, i.e. $\tilde {\boldsymbol{x}}_\epsilon = \boldsymbol{x}_\epsilon$. 
Thus $x_\epsilon^\mu$ pulled back to $\calP$ contains all necessary information about the photon position as registered by any observer.
% comment
%Czy następne zdanie jest dobrze sformułowane?
% reply - tak! equivalence classes to wektory ortogonalne i znormalizowane w P, więc baza w tej przestrzeni.
%

\paragraph{Screen vectors and orthonormal frames in $\calP$.}
We will now elaborate on the relation between the screen vectors in adapted frames and orthonormal frames in $\calP$. 
We first note that equivalence classes $\boldsymbol{e}_A$ of the screen vectors 
$e_A$ of any observer $u^\mu$ constitute an orthonormal frame in $\calP$. 

Let us now imagine another observer $v^\mu$ at the same point, with his adapted frame $(v^\mu, f\DU{A}{\mu}, s^\mu)$, who uses his own Sachs frame $f\DU{A}{\mu}$ to measure the 
position of other photons. Since $f_A \in p^\perp$, they must be related to the Sachs basis  of $u^\mu$ by 
\bea
 f\DU{A}{\mu} = R\UD{B}{A}\,e\DU{B}{\mu}  + C_A(-u^\mu+r^\mu) = R\UD{B}{A}\,e\DU{B}{\mu}  + \frac{C_A}{Q}\,p^\mu \label{eq:frote}
\eea
with yet unspecified coefficients $R\UD{A}{B}$ and $C_A$. The ortho-normality condition $f\DU{A}{\mu}\,(f_{B})_{\mu} = \delta_{AB}$ implies that $R\UD{A}{B}\,R\UD{C}{D}\,\delta_{AC} = \delta_{BD}$.  
It is also easy to prove that if both 
frames have the same orientation then $\det R\UD{A}{B} > 0$. It follows that the screen vectors must be related by an $SO(2)$ rotation
around the $r^\mu$ axis and possibly adding terms proportional to $p^\mu$. In particular, the equivalence classes $\boldsymbol{f}_A$ and $\boldsymbol{e}_A$ constitute two orthonormal frames in
$\calP$, related by a rotation:
\bea
\boldsymbol{f}_1 &=& \cos\varphi\,\boldsymbol{e}_1 + \sin\varphi\,\boldsymbol{e}_2  \nonumber\\
\boldsymbol{f}_2 &=& -\sin\varphi\,\boldsymbol{e}_1 + \cos\varphi\,\boldsymbol{e}_2,  \nonumber
\eea
Thus, the registered components of relative positions of orthogonally displaced photons can only differ by an $SO(2)$ rotation. In particular, every observer can
choose to align his or her screen vectors with the screen vectors of $u^\mu$ by a simple rotation, i.e.
\bea
\boldsymbol{e}_1 = \boldsymbol{f}_1 \label{eq:alignment}
\eea
and the same for the other basis vector.
Let us stress that
this does not mean that the whole screen vectors $e\DU{A}{\mu}$ and $f\DU{A}{\mu}$  are equal: they must be perpendicular to the 4-velocities of different observers
and will in general be different. However, they may differ only by a vector proportional to $p^\mu$, as in (\ref{eq:frote}),
so their pullbacks to $\calP$ will be the same. With this alignment, observers $u^\mu$ and $v^\mu$ will register 
exactly the same distances and angles between the positions of orthogonally  displaced photons.

The relation between the screen vectors and frames in $\calP$ implies a simple relation between the vector and tensor components. Namely, for a vector $X^\mu$ such 
that $X^\mu\,p_\mu =0$ we have 
\bea
X^A = \boldsymbol{X}^A,
\eea
where $X^A$ are the two $e\DU{A}{\mu}$ components in an adapted frame and $\boldsymbol{X}^A$ are the components of the equivalence class $\boldsymbol{X}$ in the 
corresponding orthonormal frame $\boldsymbol{e}_A$ in $\calP$. 
The reader may 
verify that identical results hold for 1-forms and for higher valence tensors perpendicular to $p^\mu$ in each index. Thus we 
can see that when performing calculations on these objects pulled back to $\calP$ we may simply focus on their $A=1,2$ components as expressed in 
an arbitrary adapted frame. These components happen to be identical to the components of the boldface equivalence class in $\calP$. 
We will make use this fact in the rest of the paper
and simply discuss the perpendicular $A=1,2$ components of $p^\mu$-perpendicular geometric objects without introducing explicitly their boldface counterparts. 
We should remember, however, that $X^A$'s can also be interpretated as components in an observer-invariant quotient space $\calP$ of reduced dimension.

The results of this section were first reported by Sachs \cite{sachs} and Ehlers, Jordan and Sachs \cite{ehlers-jordan-sachs} in the form of a 
theorem about the shadow cast on a screen by a small object placed along a null congruence emanating from a single point. 
These results, stated in the language of this paper, state that the quotient space $\calP$ contains all the information needed in the geometric optics of photons
 in $p^\perp$. Moreover, we have shown that the space $\calP$ itself and the notion of an orthonormal basis in it are observer-independent. 
More precisely, we may consider
an orthonormal basis in $\calP$ as an equivalence class of aligned Sachs bases used by all possible observers at a point. $\calP$ appeared also in differential geometry in the context of so-called optical geometries and their relation to the Cauchy-Riemann spaces \cite{robinson-trautman} . 

Finally, we note that the whole GDE can be pulled back to $\calP$. Since $\gamma_0$ is a geodesic, we see that the 
covariant derivative can be pulled back to $\calP$: let $\xi^\mu$ be a vector field along $\gamma_0$ with $\xi^\mu\,p_\mu = 0$ everywhere. Then 
\bea
\nabla_p (\xi^\mu + C(\lambda)\,p^\mu) = \nabla_p \xi^\mu + \dot C\,p^\mu,  \nonumber
\eea
i.e. the equivalence class of $\nabla_p\xi^\mu$ does not depend on the equivalence class of $\xi^\mu$. Therefore, we can consider the 
covariant derivative $\nabla_p$ as an operator acting only in $\calP$.
%comment - moze cos w stylu contracted riemnann tensor zamiast 'expression tensor'?
%reply - po prostu expression
Similarly, we note that the expression $R\UD{\mu}{\alpha\beta\nu}\,p^\alpha\,p^\beta\,\xi^\nu$ is insensitive to gauge transformations $\xi^\mu \to \xi^\mu + E\,p^\mu$, or equivalently its 
value does not depend on the choice of
the representative of $\xi^A \in \calP$. The resulting vector is perpendicular to $p^\mu$ and therefore
can be pulled back to $\calP$. Thus, the whole curvature tensor contracted twice with $p^\mu$ can be unambiguously pulled back to
$\calP$, yielding the operator $R\UD{A}{\alpha\beta B}\,p^\alpha\,p^\beta$. The equivalence class $\xi^A$ of any solution of the GDE (\ref{eq:GDEfirst}) needs to satisfy
the reduced GDE
\bea
\nabla_p \nabla_p \xi^A - R\UD{A}{\mu\nu B}\,p^\mu\,p^\nu\,\xi^B = 0. \label{eq:reducedGDE}
\eea
This is a second order ODE for 2 functions. We have reduced this way the full, 4-dimensional GDE with 8-parameter space of solutions to 
half of this dimension.

In practice, we will solve this equation by introducing an observer $u^\mu$ at a point and his adapted frame, and then propagating it in a parallel manner 
along $\gamma_0$. This way, we obtain the frame $\left(\hat u^\mu,\hat e\DU{A}{\mu},\hat r^\mu\right)$ defined along the whole null curve. It can be easily checked that the condition 
 for $\hat r^\mu$ being in the direction of propagation
 of the photon is fulfilled everywhere.
  Now the Sachs basis $\hat e\DU{A}{\mu}$ determines an orthonormal basis $\hat {\boldsymbol{e}}_A$ in $\calP$ at every point along $\gamma_0$. The components
$\xi^A$, $R\UD{A}{\alpha\beta B}\,p^\alpha\,p^\beta$, etc. denote simply the perpendicular, $\hat e\DU{A}{\mu}$ components of $\xi^\mu$ and 
$R\UD{\mu}{\alpha\beta \nu}\,p^\alpha\,p^\beta$ respectively in the adapted frame, and at the same time the components of the equivalence class in $\calP$. Equation (\ref{eq:reducedGDE}) expressed in the new frame takes a particularly simple form of
\bea
\frac{\dd^2\xi^A}{\dd \lambda^2} - R\UD{A}{\mu\nu  B}\,p^\mu\,p^\nu\,\xi^B(\lambda) = 0.  \nonumber
\eea
In analogy with (\ref{eq:GDopdef}), we introduce the reduced GDE operator acting on vector fields along $\gamma_0$ in $\calP$:
\bea
\widetilde \calG[\xi]^A = \nabla_p \nabla_p \xi^A - R\UD{A}{\mu\nu B}\,p^\mu\,p^\nu\,\xi^B.  \nonumber
\eea

\subsection{Null geodesics on a single light cone, Jacobi matrix} \label{sec:nullg}

\bfi
%\bce
\centering
%\begin{minipage}[b]{0.65\linewidth}
\includegraphics[width=0.5\textwidth]{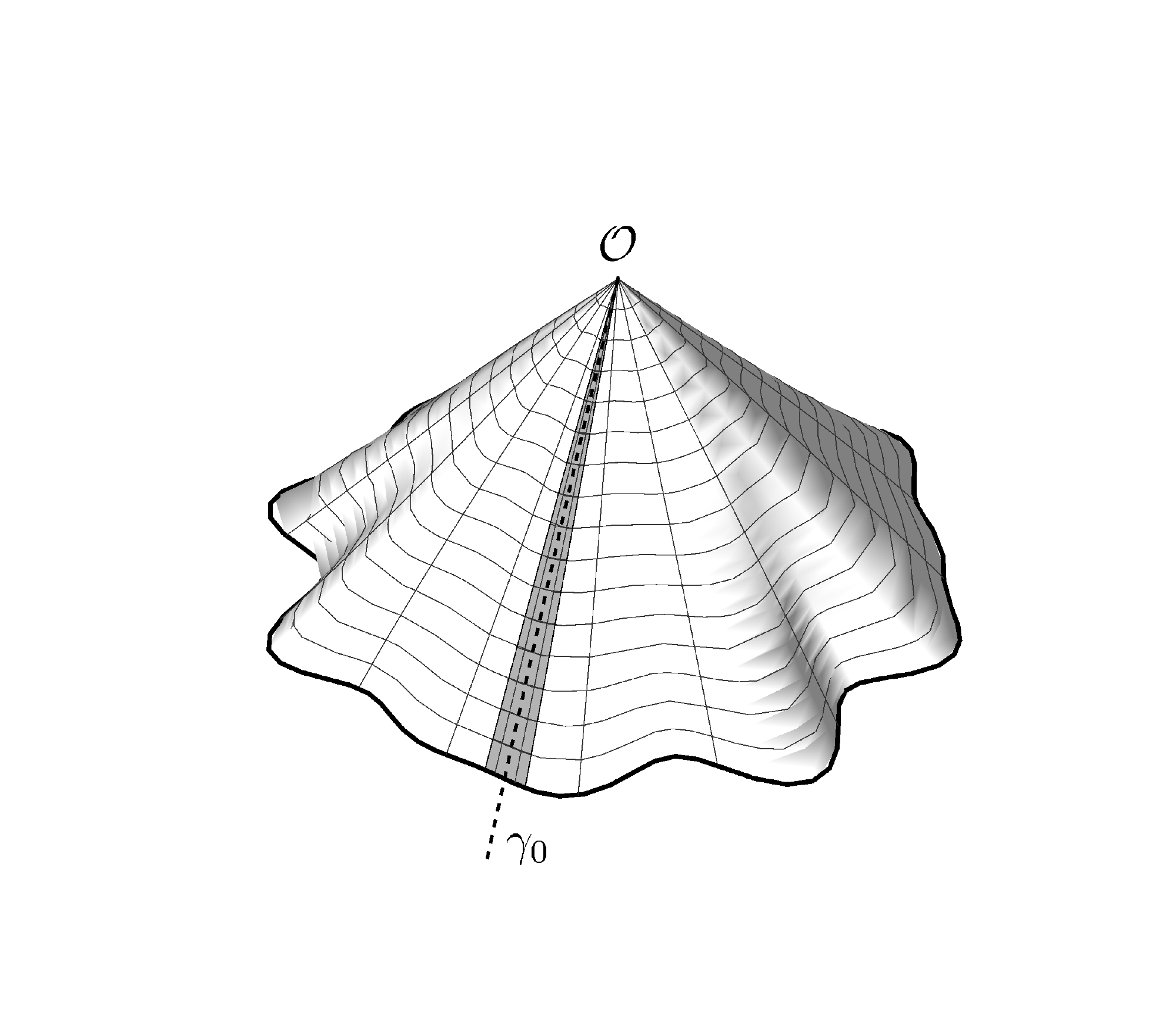}
%\end{minipage}
\caption{Thin slice of the past light cone emanating from $\calO$, formed by geodesics near $\gamma_0$. One dimension supressed.}
\label{fig:lightcone}
%\ece
\efi

In this subsection we will consider the situation when all geodesics emanate from one point $\calO$, called the observation point. This situation corresponds to considering
a thin slice of the past null cone centered at $\calO$, see Fig.~\ref{fig:lightcone}. The slice must be thin enough so that its geometry could be described by 
the first order approximation offered by the GDE. 
Let $\gamma_0$ denote now one fixed null geodesic from the light cone,
with a fixed affine parametrization
$\lambda$. 
For every neighbouring geodesic in the light cone we have
$\xi^\mu(\lambda_\calO) = 0$, where $\lambda_\calO$ corresponds to the point $\calO$ on $\gamma_0$. It follows that at $\calO$ we have $p^\mu\,\xi_\mu = 0$,
 but from 
the conservation of the product $p^\mu\,\xi_\mu$  we know that $\xi^\mu$ must be orthogonal to $p^\mu$ everywhere along $\gamma_0$. Thus, nearby photons emanating from a single point, i.e. belonging to the same light cone, are examples of photons
lying on a single wavefront. 

Note that the light cone is a 3-dimensional null surface. Thus the slice in question is also 3-dimensional. 
One dimension corresponds to the parametrization along the geodesics, while the other two label the null geodesics. The space of solutions of the GDE corresponding to
the slice must therefore be 2-dimensional, corresponding to the directions perpendicular to the light propagation.

The perpendicular part of $\xi^\mu$, i.e. $\xi^A$, satisfies the reduced GDE (\ref{eq:reducedGDE}) with the initial condition
\bea
\xi^A(\lambda_\calO) = 0.  \nonumber
\eea
The full solution requires also the first derivative $\nabla_p\xi^A$ at $\lambda_\calO$ to be specified. The solution of the reduced GDE as a function
of $\nabla_p\xi^A(\lambda_\calO)$ is given by the \emph{Jacobi matrix}:
\bea
\xi^A(\lambda) = \calD\UD{A}{B}(\lambda)\,\nabla_p\xi^B(\lambda_\calO). \label{eq:Ddef}
\eea
In an orthonormal, parallel propagated frame it can be extracted from the reduced GDE in the following way: it satisfies
\bea
\frac{\dd^2}{\dd \lambda^2}\,\calD\UD{A}{B} - R\UD{A}{\nu\alpha C}\,p^\nu\,p^\alpha\,\calD\UD{C}{B} = 0 \label{eq:JacMat}
\eea
with the initial conditions imposed at $\lambda = \lambda_\mathcal{O}$:
\bea
\calD\UD{A}{B}\left(\lambda_\mathcal{O}\right) &=& 0 \nonumber \\
\frac{\dd}{\dd \lambda}\,\calD\UD{A}{B}\left(\lambda_\mathcal{O}\right) &=& \delta\UD{A}{B}. \label{eq:initialcond} 
\eea
The ODE's above define $\calD\UD{A}{B}$ in any cooridnate system; in special coordinates, like the past null cone coordinates, there may be a simpler
way to obtain it \cite{Fanizza:2013doa, Fanizza:2014baa}.

Intuitively speaking, the Jacobi matrix describes the shape of a thin slice of the past light cone, how it is stretched, focused or defocused as the light 
propagates through the spacetime. It translates small differences in the direction of propagation of photons at a given point $\calO$ to the displacement 
in perpendicular directions with respect to the fiducial null geodesic
further on.  In our formalism it is a linear mapping taking values from $\calP$ 
at the observation point and mapping them to $\calP$ at another point $\lambda$. $\calD\UD{A}{B}$ encodes the gravitational 
lensing effects in the following way: it determines in what way a small image will be magnified or distorted in the linear approximation \cite{perlick-lrr}. 
Points at which it becomes degenerate correspond to caustics in the light propagation \cite{perlick-lrr}. 
In this paper we assume from now on that $\calD\UD{A}{B}$ is not degenerate at the emission point, i.e. the emitter is not passing through a caustic.

We stress that the Jacobi matrix $\calD\UD{A}{B}(\lambda)$, considered as a linear mapping  between the $\calP$ spaces at $\calO$ and $\gamma_0(\lambda)$
(the Jacobi operator), does not depend on the
choice of the observers at $\calO$ and $\gamma(\lambda)$. 
This is an important point, because $\calD\UD{A}{B}(\lambda)$ is 
usually defined using a parallel-propagated Sachs frame of an observer and
one might expect that its exact value depends on the value of his 4-velocity $u^\mu$. 
We see that this is not the case here, because the value of $\calD\UD{A}{B}(\lambda)$ is completely determined by the  
observer-independent \emph{equivalence classes} of the perpendicular Sachs vectors, as we have pointed out in the previous section. The components of the Jacobi matrix
recorded by various observers at $\calO$ and $\gamma_0(\lambda)$ may of course be different, but they will only differ by $SO(2)$ rotations. These
rotations can be removed
if we align the perpendicular vectors of the observers at each point as we have described above.

The Jacobi matrix $\calD\UD{A}{B}(\lambda)$ as an operator does not depend on the observer we choose at $\calO$, at point with affine parameter $\lambda$ or anywhere in between, but it does 
depend on the affine parametrization of the 
null geodesic $\gamma_0$:  it transforms under affine reparametrizations according to
\bea
\lambda &\mapsto& \lambda' = C\,\lambda + D, \qquad C,D=\const  \nonumber\\
p^\mu &\mapsto & \frac{1}{C}\,p^\mu  \nonumber\\
\calD\UD{A}{B} &\mapsto & C\,\calD\UD{A}{B},  \nonumber
\eea
because the initial conditions (\ref{eq:initialcond}) are imposed in terms of the affine parameter $\lambda$. 
 Note however that the tensor product $\calD\UD{A}{B}\,p_\mu$ is reparametrization-invariant, i.e. its value depends only on the
geometry of an infinitesimal section of a light cone.

\subsection{Redshift, angular distance and luminosity distance}

\bfi
%\bce
\centering
%\begin{minipage}[b]{0.65\linewidth}
\includegraphics[width=0.5\textwidth]{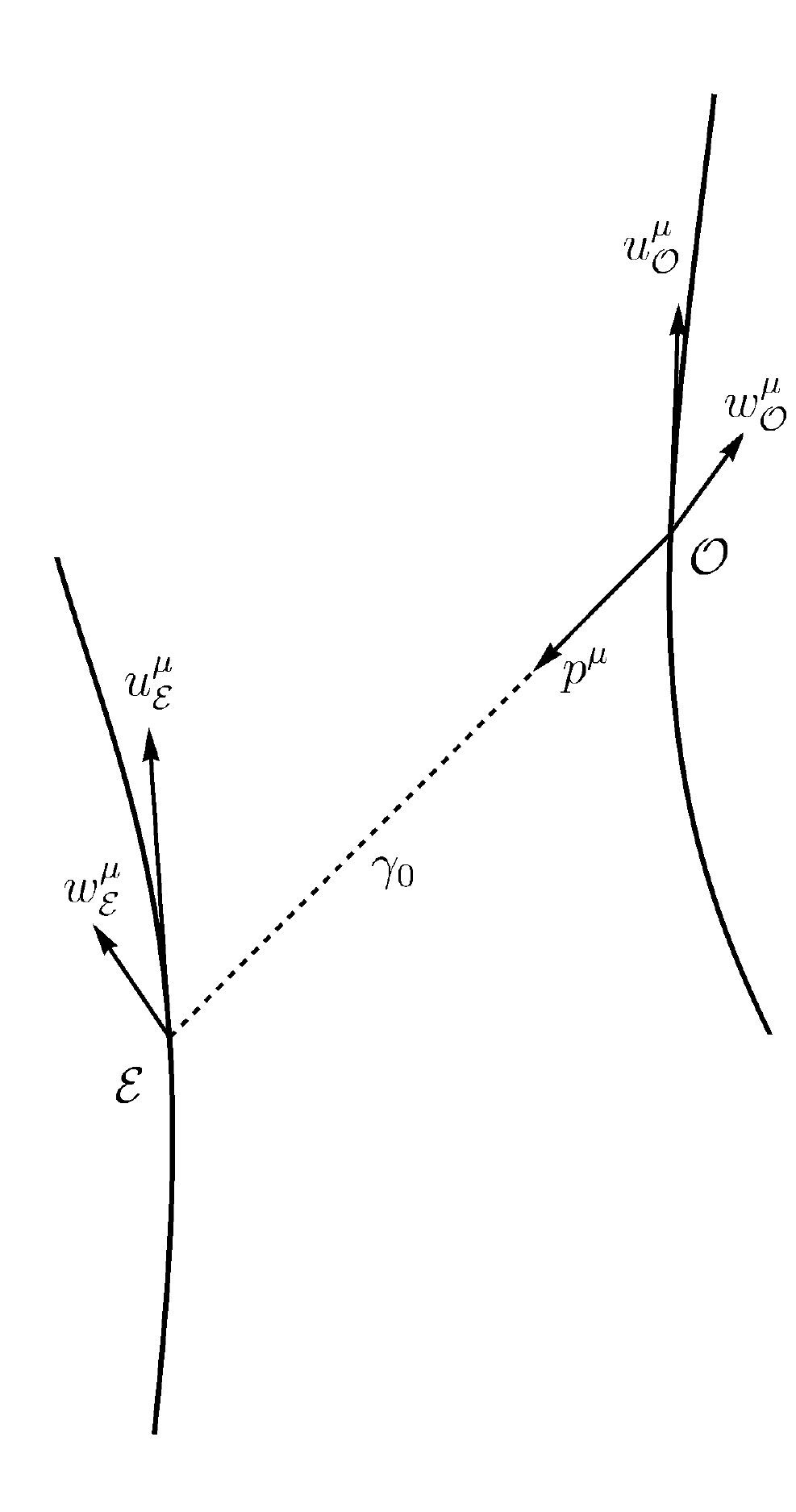}
%\end{minipage}
\caption{An emitter and an observer with a null geodesic joining their worldlines. The points of emission and observation are denoted $\calE$ and $\calO$ respectively. The
4-velocity and the 4-acceleration of the emitter in $\calE$ are denoted $u_\calE^\mu$ and $w_\calE^\mu$ respectively, and the same for the observer at $\calO$.
The tangent vector to the null geodesic is denoted $p^\mu$.}
\label{fig:situation}
%\ece
\efi

The machinery introduced in the previous section is sufficient to derive the expressions for the redshift, angular distance and luminosity distance of an emitter as viewed by
the observer. Consider now a geodesic $\gamma_0$ between an emitter with 4-velocity $u_\calE^\mu$ at a point $\calE$ and an observer $u_\calO^\mu$ at $\calO$, see 
Fig.~\ref{fig:situation}. 
The energy of the photon measured at the emitters and observers frame are proportional to $p_\mu\,u_\calE^\mu$ and $p_\mu\,u_\calO^\mu$ respectively, see (\ref{eq:sachsenergy}),
so the redshift is given by
\bea
z = \frac{p_\mu\,u_\mathcal{E}^\mu}{p_\mu\,u_\mathcal{O}^\mu} - 1 \label{eq:zdef0},
\eea
irrespective of the parametrization of $\gamma_0$.

\paragraph{Position on the observer's celestial sphere.} Let $N_\calO^-$ denote the set of null, past-oriented vectors at $\calO$. The set of directions in $N_\calO^-$, i.e.
$N_\calO^-$ divided by the relation $k^\mu \sim l^\mu$ iff $k^\mu = C\, l^\mu$ with $C>0$, is often called the \emph{celestial sphere} at $\calO$ and 
denoted  $S_\calO$ \cite{perlick, low}. We also define the \emph{sphere of directions} of the observer $\calO$, denoted by $\textrm{Dir}_\calO$, as a set of spatial vectors 
$n^\mu \in T_\calO M$,
orthogonal to the observer's 4-velocity ($n_\mu\,u_\calO^\mu = 0$) and normalized ($n^\mu\,n_\mu = 1$). Note that unlike $N_\calO^-$ and $S_\calO$, 
$\textrm{Dir}_\calO$ depends explicitly on the observer.

The observer assigns the direction from which the associated photon emanates to every past-directed null vector $l^\mu$ via the formula
\bea
l^\alpha \to n^\mu(l^\alpha) = \frac{l^\beta(\delta\UD{\mu}{\beta} + u_\calO^\mu\,u_{\calO\,\beta})}{l_\sigma\,u_\calO^\sigma}. \label{eq:celestial}
\eea
The formula above defines the mapping $W: N_\calO^{-} \to \textrm{Dir}_\calO$, which is constant on the equivalence classes in $N_\calO^{-}$ and thus
can be defined on  $S_\calO$. 
$W$ consists of an orthogonal projection to  the subspace $u_\calO^\perp$ and the normalization of the resulting vector. By substituting $p^\mu$ for $l^\mu$ 
in (\ref{eq:celestial}) we obtain the observed
position of the emitter on the observers celestial sphere, i.e. $r^\mu$ in the adapted frame of the observer.

\paragraph{Angular distance. }The angular distance of the emitter is defined by the ratio between the physical size of the emitter, as measured in its own frame,
and the angular size of the body as observed by
the observer. The cross-sectional area $A(\gamma_0)$ in the direction of $\gamma_0$ of the emitter is a natural measure of the former, while the solid angle 
occupied by the image registered
by the observer $\Omega$ may serve as the latter. In this case we have
\bea
D_{ang} = \sqrt{\frac{A(\gamma_0)}{\Omega}}.  \nonumber
\eea
We introduce the observer's and the emitter's Sachs frames, $(u^\mu_\calO,e\DU{A}{\mu},r^\mu)$ and $(u^\mu_\calE,f\DU{A}{\mu},s^\mu)$ respectively. For simplicity, we assume
the parallel propagated $\hat e\DU{A}{\mu}$ and $f\DU{A}{\mu}$ to be aligned as vectors in $\calP$ at $\calE$, as in ($\ref{eq:alignment}$). 

Let $(\widetilde{\boldsymbol{e}}^1,\widetilde{\boldsymbol{e}}^2)$, $(\widetilde{\boldsymbol{f}}^1,\widetilde{\boldsymbol{f}}^2)$ denote the co-frames in $\calP$ dual to 
$(\boldsymbol{e}_1,\boldsymbol{e}_2)$ and $(\boldsymbol{f}_1,\boldsymbol{f}_2)$ respectively.
The 
cross-sectional area can now be expressed as the integral over the $\calP$ space
at $\calE$, denoted by $\calP_\calE$, of the body cross-section $\Sigma$ with $\widetilde{\boldsymbol{f}}^1\wedge \widetilde{\boldsymbol{f}}^2$ as the area form:
\bea
A(\gamma_0) = \int_{\Sigma} \widetilde{\boldsymbol{f}}^1\wedge \widetilde{\boldsymbol{f}}^2,  \nonumber
\eea
where the cross-section $\Sigma \subset \calP_\calE$ is defined as the set of all perturbed null geodesics, emanating from $\calO$, intersecting the emitter. 
$\Sigma$ can now be mapped to $\calP_\calO$ via the Jacobi operator $\calD\UD{A}{B}(\lambda_\calE)$, yielding $\calD^{-1}(\Sigma) \subset \calP_\calO$, i.e. the set of the same
null geodesics intersecting the emitter, but now defined by the difference of their directions of propagation at the observation point.

Now we consider the size of the image of the emitter on the celestial sphere $S_\calO$, given by the mapping $W$ defined above. In the vicinity of the direction $r^\mu$, 
from which the observer registers $\gamma_0$ coming,
we can introduce a quasi-Cartesian coordinate system $(\theta^1,\theta^2)$  given by the transversal components of the vector $n^\mu \in S_\calO$, i.e.
$\theta^A = \delta^{AB}\,e\DU{B}{\mu}\,n_\mu$. It is straightforward to check that near $r^\mu$ (corresponding to $\theta^A = 0$)
the solid angle element is simply $\dd\theta^1\wedge\dd\theta^2 + 
O\left(\left|\theta\right|^2\right)$. We also linearize $W$ in the vicinity of $p^\mu$:
\bea
W^\mu(p^\sigma + \delta p^\sigma) = r^\mu + \frac{\delta p^\sigma\left(\delta\UD{\mu}{\sigma} + (u_\calO^\mu - r^\mu)\,u_{\calO\,\sigma}
\right)}{p_\kappa\,u_\calO^\kappa} + O(|\delta p|^2).  \nonumber
\eea
We restrict the formula above to null vectors, i.e. to perturbations satisfying $\delta p^\sigma\,p_\sigma$. In this case $\delta p^\mu \, r_\mu = \delta p^\mu \, u_{\calO\,\mu}$, so
\bea
W^\mu(p^\sigma + \delta p^\sigma) = r^\mu + \frac{\delta p^\sigma\left(\delta\UD{\mu}{\sigma} 
+ u_\calO^\mu\,u_{\calO\,\sigma} - r^\mu\,r_\sigma\right)}{p_\kappa\,u_\calO^\kappa} + O(|\delta p|^2).  \nonumber
\eea
The linear part is thus the normalized orthogonal projection of $\delta p^\mu$ to the 2-dimensional screen subspace, orthogonal to both the direction of propagation and the
observer's 4-velocity. In the $\theta^A$ coordinates this means that
\bea
\theta^A\left(W(p^\mu+\delta p^\mu)\right) = \frac{\delta p^A}{p_\kappa\,u_\calO^\kappa} + O(|\delta p|^2), \label{eq:theta}
\eea
where $\delta p^A$ denotes the transversal components in the Sachs frame. As we noted, if $\delta p^\mu \in \calP$, then $\delta p^A$ is also equal to the components in the
corresponding $\boldsymbol{e}_A$ basis in $\calP$, so we note that the pullbacks near $p^\mu$ satisfy
\bea
W^*(\dd\theta^1) = \frac{1}{p_\kappa\,u_\calO^\kappa}\,\widetilde{\boldsymbol{e}}^1 &+& O(|\delta p|)  \nonumber\\
W^*(\dd\theta^2) = \frac{1}{p_\kappa\,u_\calO^\kappa}\,\widetilde{\boldsymbol{e}}^2 &+& O(|\delta p|).  \nonumber
\eea

We are now ready to evaluate the solid angle $\Omega$:
\bea
\Omega = \int_{W\circ\calD^{-1}(\Sigma)} \dd\theta^1\wedge\dd\theta^2 = \frac{1}{\left(p_\kappa\,u_\calO^\kappa\right)^2}\int_{\calD^{-1}(\Sigma)} \widetilde{\boldsymbol{e}}^1\wedge \widetilde{\boldsymbol{e}}^2  \nonumber
\eea
to linear order in the 4-momentum deviation. The last expression can be easily related to the integral at the emission point if we recall that 
the Jacobi operator $\calD$ is linear and given in these frames by the Jacobi matrix $\calD\UD{A}{B}(\lambda_\calE)$:
\bea
\Omega = \frac{1}{\left|\det \calD\UD{A}{B}(\lambda_\calE)\right|\,\left(p_\kappa\,u_\calO^\kappa\right)^2} \int_{\Sigma} \widetilde{\boldsymbol{f}}^1\wedge \widetilde{\boldsymbol{f}}^2. \nonumber
\eea
We obtain immediately the standard expression for the angular distance
\bea
D_{ang}=\left(p_\mu\,u_\mathcal{O}^\mu\right)\,\left| \det \calD\UD{A}{B}\left(\lambda_\mathcal{E}\right) \right|^{1/2} \label{eq:Dangdef}
\eea
independent of the emitting body's shape $\Sigma$. 

\paragraph{ Luminosity distance. } The (uncorrected) luminosity distance is defined in an analogous manner, using the ratio of the observed  luminosity $L$ and
the energy flux $F$ of the emitter's radiation measured by the observer
\bea
 D_{lum} = \sqrt{\frac{L}{4\pi F}},  \nonumber
\eea
%comment - moze zacytowac wprost ta prace Etheringtona?
%reply - tak, i pewnie coś jeszcze.
see \cite{etherington, etherington2, perlick-lrr}. It can be calculated using a similar approach with the role of the observer and emitter reversed and considering the energy flux of the 
photons, but it is a classical result by Etherington \cite{etherington, etherington2} that it is related to $D_{ang}$ and $z$ via the distance duality relation
\bea
D_{lum} = D_{ang} (1+z)^2 \label{eq:etherington}.
\eea

\subsection{Dependence on the observer, emitter and spacetime} \label{sec:dependence1}

Let us now think how the optical observables  $z$, $D_{ang}$ and $D_{lum}$ depend on the emitter, observer and the geometry of the spacetime.

The redshift obviously depends on both the emitter's and the observer's 4-velocities and positions:
\bea
z \equiv z(u_\calO^\mu,u_\calE^\mu).  \nonumber
\eea
Indirectly, it also depends on the geometry of spacetime, since the parallel transport and the null geodesic depend on it.

The angular distance given by (\ref{eq:Dangdef}) obviously depends on the Riemann tensor along the null geodesic,
since we need it to obtain the Jacobi matrix via equation (\ref{eq:JacMat}). It also depends on the 4-velocity of the observer, but rather 
surprisingly \emph{not} on the emitter's 4-velocity:  
\bea
D_{ang} \equiv D_{ang}(R\UD{\mu}{\nu\alpha\beta},u_\calO^\mu).  \nonumber
\eea
The
dependence on the observer is only via the product $p_\mu\,u^\mu_\calO$. The product $|\det \calD\UD{A}{B}(\lambda_\calE)|^{1/2}\,p_\mu$ is parametrization-independent
 and independent of the observer's 4-velocity. As we noted in Sec. \ref{sec:nullg}, it depends only on the geometry of the past light cone centered at $\calO$.

 The geometric reason for this rather surprising independence of $u_\calE^\mu$ can be traced back to the results of Section \ref{sec:nullgde}, i.e. 
 possibility of pulling back the 
 GDE to the observer-independent space $\calP$. Recall that in $\calP$ 
 the distances between photons, as measured on screen spaces of different observers, are the same. $D_{ang}$ is a function of the 
 projection of separation vectors for solutions with a common origin at $\calO$ to the screen space of the emitter. We have shown in Section \ref{sec:nullgde} that this projection is independent of the 4-velocity defining the screen space (up to irrelevant rotations) at any point because the vectors in question are perpendicular to $p^\mu$. 
 Thus, there is no net dependence on $u_\calE^\mu$ in $D_{ang}$. On the other hand, the luminosity distance defined by (\ref{eq:etherington}) depends on the curvature and  both 4-velocities:
\bea
D_{lum} = D_{lum}(R\UD{\mu}{\nu\alpha\beta},u_\calO^\mu,u_\calE^\mu).  \nonumber
\eea

\section{Drift effects}

We will now add a time dimension to the geometric optics formalism in GR and ask about the position, redshift and Jacobi matrix drift measured by the observer  
given the 4-velocities and 4-accelerations of the emitter and the observer.

\subsection{Geometry}
\bfi
%\bce
\centering
%\begin{minipage}[b]{0.65\linewidth}
\includegraphics[width=0.5\textwidth]{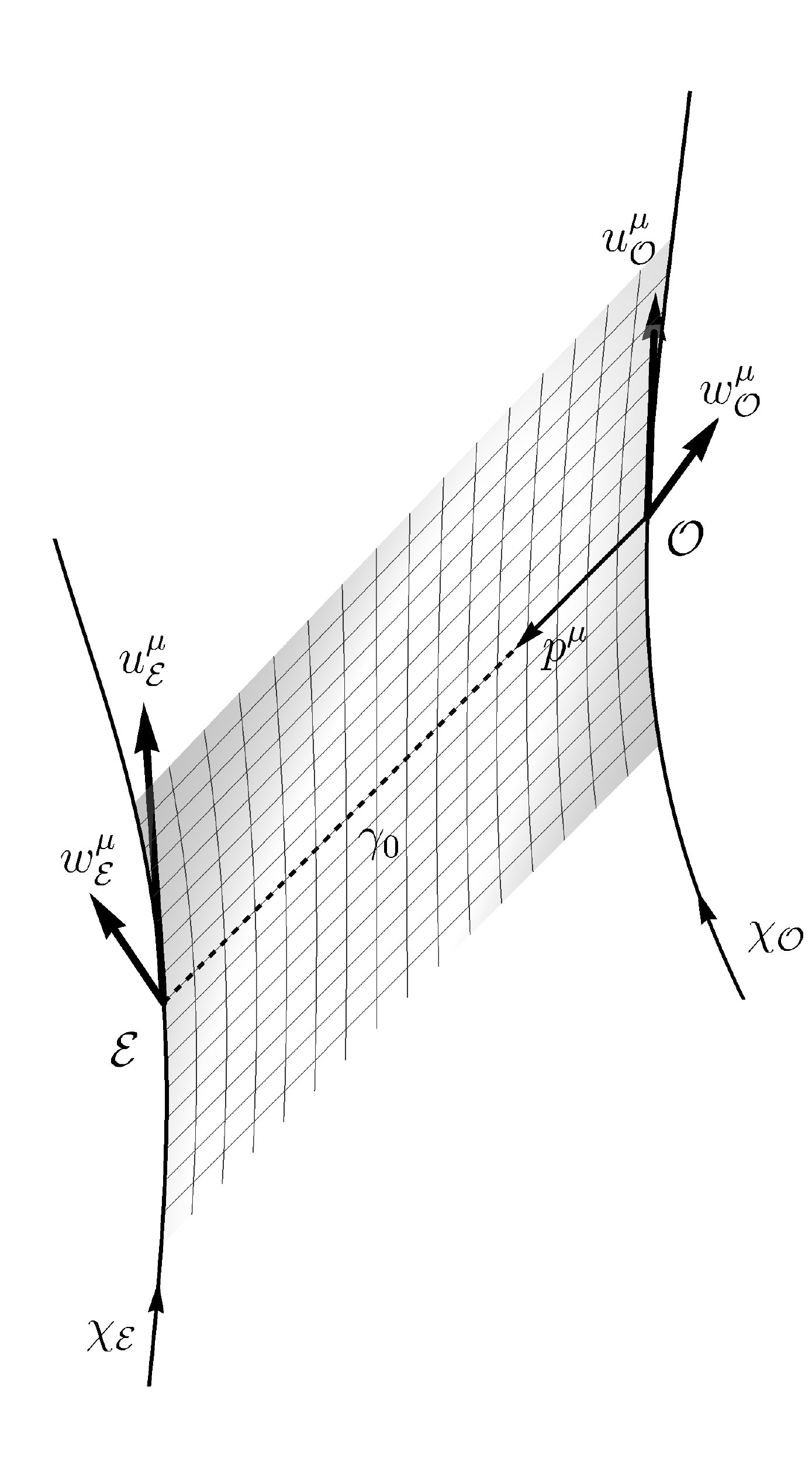}
%\end{minipage}
\caption{The worldlines $\chi_\calE$ and $\chi_\calO$ of the emitter and the observer respectively. The null surface spanned between them consists of null geodesics 
$\gamma_\tau$ connecting points on both worldlines. In each connected pair of points, the points on $\chi_\calO$ lie in the causal future of those from $\chi_\calE$. 
$\gamma_0$ connects points $\calO$ and $\calE$. The 4-velocity and 4-acceleration of the emitting body at $\calE$ is $u^\mu_\calE$ and $w^\mu_\calE$ respectively.
The 4-velocity and 4-acceleration of the observer at $\calO$ is $u^\mu_\calO$ and $w^\mu_\calO$ respectively. The null, past-oriented vector $p^\mu$ is the tangent vector
to each $\gamma_\tau$.}
%An observer with 4-velocity $u^\mu$ along the null geodesic $\gamma_0$, the backward-pointing 4-momentum of the photon $p^\mu$,
%the 3 remaining Sachs frame vectors: $r^\mu$, pointing in the direction from which the photon is coming and the screen vectors $e\DU{A}{\mu}$. %The spatial
%3-space $u_\perp$ denotes the plane of simultaneity of the observer.

\label{fig:driftsituation}
%\ece
\efi

Consider now the situation from Fig.~\ref{fig:driftsituation}. The observer and the emitter move along their worldlines $\chi_\calO$ and $\chi_\calE$. The worldlines are connected by a family of null 
geodesics $\gamma$, i.e. there is exactly one null geodesic from any point along $\chi_\calO$, past-oriented and ending in a single point along $\chi_\calE$. We will 
now make gauge choices concerning the parametrization of the curves. 
$\chi_\calO$ will be parametrized by the observer's proper time $\tau$, $\chi_\calE$ by the emitter's proper time $\sigma$ and the family of null geodesics by the 
proper time $\tau$ in which a given $\gamma$ intersects $\chi_\calO$. This way we parametrize the family by the observation time of a given photon registered by the observer.
On the other hand, we fix the affine parametrization $\lambda$ of each null geodesic $\gamma_\tau$ by demanding that intersection with the observer and emitter took place for
fixed values $\lambda_\calO$ and $\lambda_\calE$ of the parameter $\lambda$. An affine reparametrization can always ensure that and
these conditions fixes the parametrization of the null geodesics family up to the irrelevant choice of the
$\tau=0$ moment on the observer's worldline. With this setup we have
%comment - co to jest s(t) w ponizszym wzorze?
%reply - dodane
\bea
 \gamma_\tau(\lambda_\calO) &=& \chi_\calO(\tau) \label{eq:congruencecond1}\\
 \gamma_\tau(\lambda_\calE) &=& \chi_\calE\left(s(\tau)\right) \label{eq:congruencecond2}.
\eea
$s(\tau)$ gives the relation between the proper time of the observer carried by the light signals to the emitter and the emitter's proper time.
Consider now the null geodesic $\gamma_\tau$ for a fixed $\tau=\tau_0$, connecting the photon emission at point $\calE \equiv \chi_\calE(\sigma_0)$ with the observation at $\calO \equiv
\chi_\calO(\tau_0)$, denoted as before by $\gamma_0$. We 
ask about the deviation vector $X^\mu$ corresponding to the null geodesic 
registered at an infinitesimally later moment $\tau_0 + \dd\tau$ by the observer. We will call $X^\mu$ the \emph{observation time vector}, see Fig.~\ref{fig:Xvector}. 
It turns out it can be expressed via the 4-velocities $u^\mu_\calO$ and $u^\mu_\calE$. 
%comment- w ponizszym obrazku dwie ostatnie strzalki po prawej stronie sa troche krzywo
%reply - poprawię wkrótce
\bfi
%\bce
\centering
%\begin{minipage}[b]{0.65\linewidth}
\includegraphics[width=0.5\textwidth]{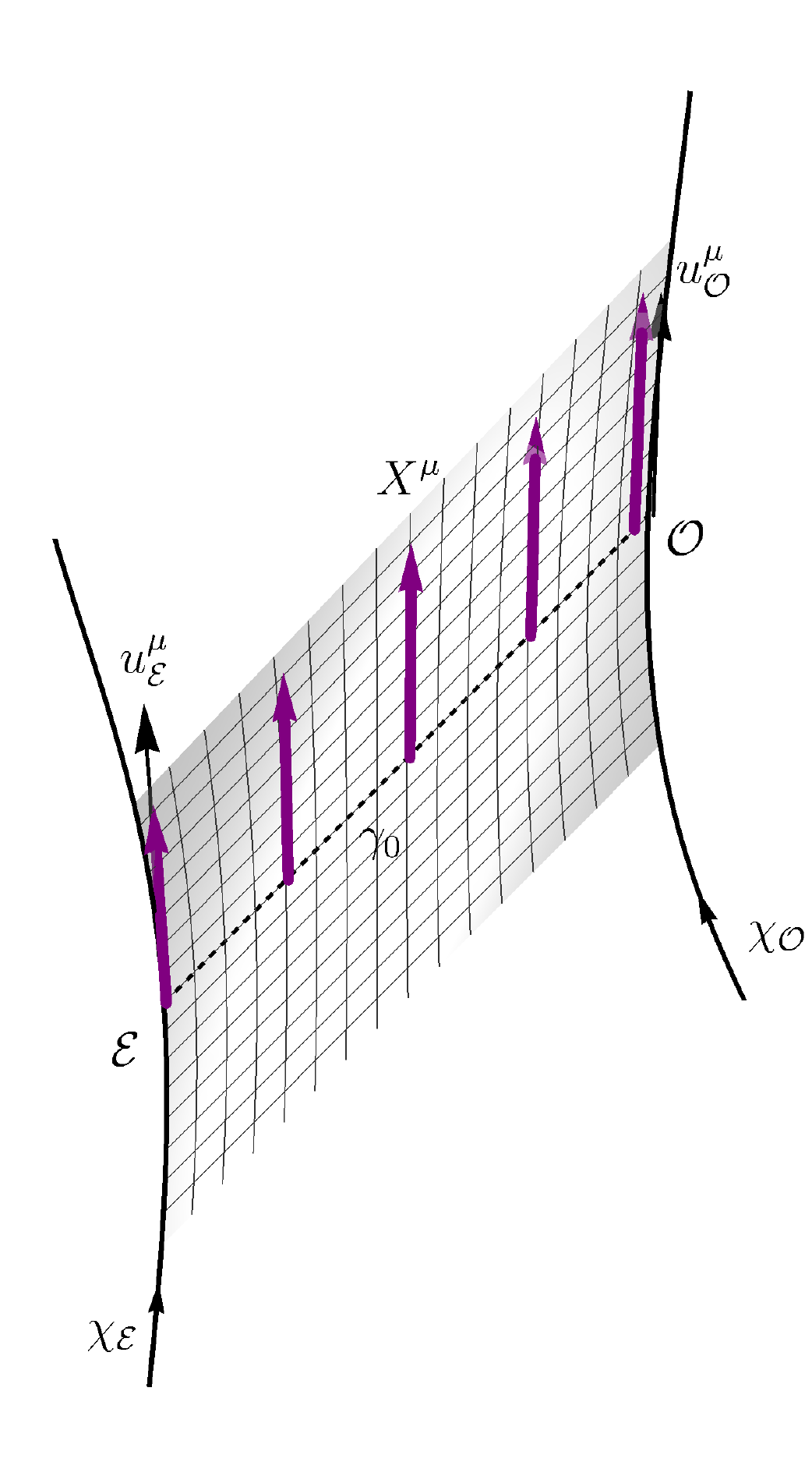}
%\end{minipage}
\caption{The observation time vector points to the null geodesic at an infinitesimally later moment as measured by the observer. It satisfies the GDE and 
interpolates between $u_\calO^\mu$ at $\cal O$ and $\frac{1}{1+z}\,u_\calE^\mu$ at $\calE$.}
%An observer with 4-velocity $u^\mu$ along the null geodesic $\gamma_0$, the backward-pointing 4-momentum of the photon $p^\mu$,
%the 3 remaining Sachs frame vectors: $r^\mu$, pointing in the direction from which the photon is coming and the screen vectors $e\DU{A}{\mu}$. %The spatial
%3-space $u_\perp$ denotes the plane of simultaneity of the observer.

\label{fig:Xvector}
%\ece
\efi

The observation time vector satisfies the GDE 
\bea
\nabla_p\nabla_p X^\mu &=& R\UD{\mu}{\nu\alpha\beta}\,p^\nu\,p^\alpha\,X^\beta \label{eq:XGDE}
\eea
with the following conditions:
\begin{enumerate}
 \item at $\calO$ it needs to coincide with the 4-velocity of the observer $u^\mu_\calO$, i.e.
 \bea
  X^\mu(\lambda_\calO) = u^\mu_\calO, \label{eq:XboundO}
 \eea 
 \item at $\calE$ it needs to be proportional to the 4-velocity of the observer $u^\mu_\calO$, i.e.
 \bea
  X^\mu(\lambda_\calE) = C\,u^\mu_\calE \label{eq:XboundE}
 \eea 
 for a positive $C>0$,
 \item it has to correspond to a family of null vectors, so from (\ref{eq:null2}) we have
 \bea
  (\nabla_p X^\mu)\,p_\mu = 0 \label{eq:Xnullcond}
 \eea
 imposed anywhere along $\gamma_\tau$.
\end{enumerate}
Conditions 1 and 2 follow directly from (\ref{eq:congruencecond1}--\ref{eq:congruencecond2}), which in turn follow from
the parametrization condition we have imposed on the family $\gamma_\tau$. They can be interpreted as follows: if the meeting points of 
all null geodesics with the
observer's and emitter's worldlines correspond to the same values $\lambda_\calO$ and $\lambda_\calE$ of the affine parameter, then the vector $X^\mu$, corresponding to 
an infinitesimal change of the parameter $\tau$ without affecting $\lambda$, must be aligned  along the worldlines $\chi_\calO$ and $\chi_\calE$ at
$\calO$ and $\calE$ respectively. Moreover, since the parametrization of the family $\gamma_\tau$ is supposed to agree with the proper time along 
the observer's worldline, $X^\mu$ must in fact be exactly equal to the tangent vector to $\chi_\calO$.

The constant $C$ is a priori unknown, but it turns out it can be obtained from Condition 3. As we have noted in Sec.~\ref{sec:nullgde}, (\ref{eq:Xnullcond})
is equivalent to imposing the condition $p_\mu \,X^\mu$ to be equal at 2 different points. If we take them to be $\calO$ and $\calE$
and make use of (\ref{eq:XboundO}--\ref{eq:XboundE}), we obtain $p_\mu\,u^\mu_\calO = C\,p_\mu\,u^\mu_\calE$
or $C = (1 + z)^{-1}$ with $z$ being the redshift. Thus the conditions above can be reduced to 
\bea
 X^\mu(\lambda_\calO) &=& u^\mu_\calO \label{eq:XboundO1}\\
 X^\mu(\lambda_\calE) &=& \frac{1}{1+z}\,u^\mu_\calE,  \label{eq:XboundE1}
\eea
with $z$ given by the standard redshift formula (\ref{eq:zdef0}). 

The coefficient $C$ has the physical interpretation of the ratio between the emitter's proper time lapse between two close points $\chi_\calE(\sigma)$ and 
$\chi_\calE(\sigma+\dd\sigma)$, positioned along the worldline of the emitter in the vicinity of $\calE$, and the time lapse measured 
by the observer between registering the emission from each of them. Thus  we have confirmed the classical result stating that if the observer sees the 
photons from $\calE$ redshifted by $z$, he or she will also see the emitter's proper time lapse slowed down by the factor of $1+z$, see \cite{perlick, kermack_mccrea_whittaker_1934}.

The observation time vector $X^\mu$ is the main geometric tool we will use in this paper. It has already appeared in \cite{perlick} under the name of 
``infinitesimal message''. Its significance  stems from the fact that the covariant derivative $\nabla_X$, applied to $r^\mu$, $z$ or $\calD\UD{A}{B}$, yields
the position, redshift and Jacobi matrix drifts respectively. Conditions (\ref{eq:XboundO1}--\ref{eq:XboundE1}) constitute a boundary value problem for the 2nd order ODE (\ref{eq:XGDE}), sufficient to yield
a single solution in the generic case. In practice, it should be possible to solve
using numerical methods, but as we will see in the next section it can be solved almost completely with the help of the Jacobi matrix formalism.

\subsection{Observation time vector from the Jacobi matrix} \label{sec:Xvect}
% CHANGED SINCE SUBMISSION

The null geodesics corresponding to $X^\mu$ are not displaced orthogonally with respect to $\gamma_0$, since $p_\mu\,X^\mu = p_\mu\,u_\calO^\mu \neq 0$, but we may nevertheless isolate the orthogonal component of $X^\mu$.
Define $\hat u_\calO^\mu$ as the parallel transport of the observer's 4-velocity along $\gamma_0$:
\bea
 \nabla_p \hat u_\calO^\mu &=& 0 \label{eq:uparallel}\\
 \hat u_\calO^\mu(\lambda_\calO) &=& u_\calO^\mu \label{eq:uparallelID}.
\eea
(In a parallel propagated adapted frame $\hat u_\calO^\mu$ is already available.)
Introduce now a new variable $b^\mu$  defined by
\bea
 X^\mu = \hat u_\calO^\mu + b^\mu.  \nonumber
\eea
The new variable  needs to satisfy an inhomogeneous GDE in the form of
\bea
\calG[b]^\mu = R\UD{\mu}{\alpha\beta\nu}\,p^\alpha\,p^\beta\,\hat u_\calO^\nu, \label{eq:bGDE}
\eea
where the geodesic deviation operator $\calG$ is defined by (\ref{eq:GDopdef}), and
with initial data
\bea
 b^\mu(\lambda_\calO) &=& 0 \label{eq:bboundary1}\\
 b^\mu(\lambda_\calE) &=& \frac{1}{1+z}\,u_\calE^\mu - \hat u^\mu_\calO. \label{eq:bboundary2}
\eea
The advantage of this change of variables is two-fold. Firstly, the new variable is automatically orthogonal to  $p_\mu$, therefore if we are
not interested in the $p^\mu$ component of the observation time vector $X^\mu$, we may simply solve (\ref{eq:bGDE}) in the $\calP$ space.
Secondly, note that the initial condition (\ref{eq:bboundary1}) calls for vanishing of $b^\mu$ at $\calO$,  fully consistent with the definition of the Jacobi 
matrix (\ref{eq:Ddef}). This opens up
the possibility to use $\calD\UD{A}{B}(\lambda)$ to obtain an expression for $b^A$ and consequently also $X^\mu$, but in order to do it we need to make one more step.
Making again use of linearity of the GDE, we split $b^\mu$ into the solution of the homogeneous equation with the  boundary data (\ref{eq:bboundary1}-\ref{eq:bboundary2}) 
and the solution of the inhomogeneous
equations with trivial initial conditions:
\bea
b^\mu = \phi^\mu + m^\mu,  \nonumber
\eea
where $m^\mu$ solves the inhomogeneous GDE with a simple initial condition
\bea
\calG[m]^\mu &=& R\UD{\mu}{\alpha\beta\nu}\,p^\alpha\,p^\beta\,\hat u_\calO^\nu \label{eq:mGDE}\\
m^\mu(\lambda_\calO) &=& 0  \nonumber \\
\nabla_p m^\mu(\lambda_\calO) &=& 0.  \nonumber
\eea
 The correction term $\phi^\mu$ needs to ensure that $b^\mu$ satisfies the boundary condition (\ref{eq:bboundary1}--\ref{eq:bboundary2}). It is straightforward to 
 check that it has to satisfy the homogeneous GDE with 
 boundary conditions:
 \bea
  \calG[\phi]^\mu &=&0  \nonumber\\
  \phi^\mu(\lambda_\calO) &=& 0  \nonumber\\
  \phi^\mu(\lambda_\calE) &=& \frac{1}{1+z}u_\calE^\mu - \hat u_\calO^\mu - m^\mu. \label{eq:phiboundary2}
 \eea
Note that both terms on the right hand side of (\ref{eq:phiboundary2}), i.e. $m^\mu(\lambda_\calE)$ and the 4-velocity difference  $\frac{1}{1+z}u_\calE^\mu - \hat u_\calO^\mu$, are 
orthogonal to $p_\mu$, even though each of the 4-velocities is not.

As it turns out, in order to calculate the position and redshift drifts we only need to know $X^\mu$ up to terms proportional to 
 the null vector $p^\mu$. In this case it is enough to consider (\ref{eq:mGDE}--\ref{eq:phiboundary2}) in the quotient space $\calP$:
\bea
\widetilde\calG[m]^A &=& R\UD{A}{\nu\alpha\beta}\,p^\nu\,p^\alpha\,\hat u_\mathcal{O}^\beta \label{eq:mGDEP}\\
m^A\left(\lambda_\mathcal{O}\right) &=& 0 \label{eq:mID1P}\\
\nabla_p m^A\left(\lambda_\mathcal{O}\right) &=& 0 \label{eq:mID2P}\\
\widetilde\calG[\phi]^A \label{eq:phiGDEP} &=& 0 \\
\phi^A\left(\lambda_\mathcal{O}\right) \label{eq:phiboundary1P} &=& 0 \\
 \phi^A\left(\lambda_\mathcal{E}\right) &=& \left(\frac{1}{1+z}\,u_\mathcal{E} - \hat u_\mathcal{O} \right)^A - m^A. \label{eq:phiboundary2P}
\eea 
 The first term on the rhs of (\ref{eq:phiboundary2P}) is the 4-velocity difference pulled back to $\calP$. The expression $\frac{1}{1+z}\,u_\mathcal{E}^\mu - \hat u_{\cal O}^\mu$ is automatically orthogonal to $p_\mu$. Its pullback to $\calP$ can be interpreted as the transverse component of corrected 4-velocity difference between the emitter and the observer.
Equations (\ref{eq:phiGDEP}-\ref{eq:phiboundary2P}) can be solved explicitly with the help of the Jacobi matrix (\ref{eq:Ddef}). 
Note first that $\nabla_p \phi^A(\lambda_\calO)$ is equal to $\nabla_p X^A(\lambda_\calO)$ and that it is related to the value of $\phi^A$ at $\calE$ via
$\phi^A(\lambda_\calE) = \calD\UD{A}{B}(\lambda_\calE)\,\nabla_p \phi^B(\lambda_\calO)$, so
\bea
\nabla_p \phi^A\left(\lambda_\mathcal{O}\right) = \nabla_p X^A(\lambda_\calO) = 
{\mathcal{D}^{-1}}\UD{A}{B}\left(\lambda_\mathcal{E}\right)\,\left( \left(\frac{1}{1+z}\,u_\mathcal{E} - \hat u_\mathcal{O} \right)^B - m^B \right).
\label{eq:dphiA}
\eea
The value of the derivative at $\calO$ allows now to express the whole solution along $\gamma_0$
\bea
\phi^A\left(\lambda\right) =\calD\UD{A}{B}(\lambda)\, 
{\mathcal{D}^{-1}}\UD{B}{C}\left(\lambda_\mathcal{E}\right)\,\left( \left(\frac{1}{1+z}\,u_\mathcal{E} - \hat u_\mathcal{O} \right)^C - m^C \right).
\label{eq:phiwhole}
\eea

Summarizing, $X^\mu$ is the sum of 3 terms: 
\bea
X^\mu = \hat u_\calO^\mu + m^\mu + \phi^\mu.  \label{eq:totalX}
\eea
They can be calculated (up to terms of type $C\,p^\mu$) by first transporting $u_\calO^\mu$ in a parallel manner along $\gamma_0$ according to 
(\ref{eq:uparallel}-\ref{eq:uparallelID}), 
then solving (\ref{eq:mGDEP}-\ref{eq:mID2P}) for the vector $m^A$ and using (\ref{eq:phiwhole}) to obtain $\phi^A$. 
% CHANGED SINCE SUBMISSION

\subsection{Position drift}
We show now that the drift of the position of the emitter observed by the observer, evaluated with respect to the Fermi-Walker derivative, can be expressed using the value
of the derivative $\nabla_p\phi^A$ from (\ref{eq:dphiA}). Recall first that the Fermi-Walker derivative of a tensor defined along a (not necessary geodesic) timelike curve, with
the normalized tangent vector $u^\mu$ and 4-acceleration $w^\mu$, is defined
as
\bea
\delta_u T\UD{\alpha\dots}{\beta\dots} &=& \nabla_u  T\UD{\alpha\dots}{\beta\dots} + T\UD{\mu\dots}{\beta\dots}\,F\UD{\alpha}{\mu} + \cdots  \nonumber \\
&&-  T\UD{\alpha\dots}{\mu\dots}\,F\UD{\mu}{\beta} - \cdots  \nonumber\\
F\UD{\mu}{\nu} &=&  -u^\mu\,w_\nu + w^\mu\,u_\nu. \label{eq:FWdef}
\eea
In a geodesic it coincides with the standard covariant derivative. It is straightforward to prove its elementary properties:
\begin{enumerate}
 \item it preserves the metric tensor and thus the scalar product: $\du g_{\mu\nu} = 0$,
 \item the derivative of $u^\mu$ vanishes identically, i.e. $\du u^\mu = 0$,
 \item if a vector $\xi^\mu$ is perpendicular to $u^\mu$ ($\xi^\mu\,u_\mu = 0$), then so is its derivative, i.e. $\left(\du\xi\right)^\mu\, u_\mu = 0$.
\end{enumerate}
The Fermi-Walker transport of the orthonormal frame associated with the observer $u^\mu$, consisting of $u^\mu$ itself and 3 spatial vectors $e_i^\mu$, 
is defined by the condition $\du e_i^\mu = 0$. It corresponds to dragging the spatial vectors along the curve in a way that reacts to the change of the
inertial frame due to the observer's 4-acceleration, but otherwise admits no rotation of the spatial vectors. 

Let $\doo$ denote the Fermi-Walker derivative with respect to the observer's worldline.  
Consider now $\doo r^\mu$, i.e. the Fermi-Walker derivative of the vector $r^\mu$ from (\ref{eq:sachsrE}), as defined by the observer $\calO$ and his adapted frame. 
Since $r^\mu$ is by definition perpendicular to $u_\calO^\mu$, then so is $\doo r^\mu$. $r^\mu$ is also normalized to unity, so
$r_\mu\,\doo r^\mu = 0$. Thus $\doo r^\mu$ is orthogonal to both $u_\calO^\mu$ and $r^\mu$, and consequently also to $p^\mu$. It lies therefore in 
the observer's screen space, spanned by $e\DU{A}{\mu}$.  The definition (\ref{eq:FWdef}) yields:
\bea
\doo r^\mu = \nabla_X r^\mu - u_\calO^\mu\,w_\calO^\alpha\,r_\alpha. \nonumber
\eea
The second term is irrelevant, as it is perpendicular to the observer's screen space and will need to cancel eventually. The first one is
\bea
\nabla_X r^\mu = -\frac{\dot Q}{Q^2}\,p^\mu + Q^{-1}\,\nabla_p X^\mu + \nabla_X u^\mu. \nonumber
\eea
Again, the first term is perpendicular to the screen space, so only the second and the third ones contribute. Since $X^\mu = u_\calO^\mu$ at $\calO$, the third one is
simply the observer's 4-acceleration, while the second one can be simplified using the commutation relation (\ref{eq:commutation}) and (\ref{eq:dphiA}). The final result is
\bea
\doo r^A = \frac{1}{p_\sigma\,u_{\cal O}^\sigma}\,
{\cal D}^{-1}(\lambda_{\cal E})\UD{A}{B}\left(\left(\frac{1}{1+z}\,u_{\cal E} - \hat u_{\cal O}\right)^B - m^B\right) +  w_{\cal O}^A. \label{eq:posdrift}
\eea
This formula relates the  position drift to the observer's local 4-acceleration projected to the plane perpendicular to the line of sight (light aberration
effect), the Jacobi matrix representing the image distortion due to the lensing effects, the perpendicular component of the 4-velocity difference 
between the emitter and the observer, evaluated using the parallel transport along $\gamma_0$, as well as
a solution to the inhomogeneous GDE along the null geodesic. 
The expression $ \frac{1}{p_\sigma\,u_{\cal O}^\sigma}\,{\cal D}^{-1}(\lambda_{\cal E})\UD{A}{B}$ corresponds exactly to the lensing as seen
in the observer's frame irrespective of the null geodesic parametrization. 
Note also that since $\delta_\calO$ preserves the metric and the subspace orthogonal to $u_\calO^\mu$, it also preserves the angular distances between fixed directions
on the celestial sphere and thus the right hand side of
(\ref{eq:posdrift}) can be directly related to the rate of change of angular distance between two sources on the observer's sky. 

Equation (\ref{eq:posdrift}) shows that strong lensing between $\calO$ and $\calE$ may enhance or reduce the position drift, depending on the shape of the 
image distortion and the direction of motion of both emitter and observer. In particular, if the source is located at a caustic we may have formally $\delta_\calO r^A \to \infty$.
In practice the drift will be limited by the finite resolution of the observer's instruments and the finite size of the emitter,  like in the case of gravitational microlensing.

Note also that $m^A$ is likely to be small for short distances because
of the initial data (\ref{eq:mID1P}--\ref{eq:mID2P}), so the drift will be dominated by the first term, proportional to the perpendicular component of the difference 
between the 4-velocities 
of the observer and emitter,  
evaluated using parallel propagation along $\gamma_0$. On the other hand the $u_\calE^\mu$-independent  term $m^A$ may turn out to be 
relevant on cosmological scales.  

The $m^A$ term as defined by (\ref{eq:mGDEP}-\ref{eq:mID2P}) is a linear function of $u_\calO^\mu$ and can therefore be written in the form of
\bea
m^A = m\UD{A}{\mu}(\lambda_\calE)\,u_\calO^\mu,
\eea
where $m\UD{A}{\mu}(\lambda)$ is a linear operator from the tangent space at $\calO$ to $\calP$ at $\gamma_0(\lambda)$. Just like the Jacobi matrix, it is a property 
of the spacetime between points $\calO$ and $\calE$, independent of the observer's and emitter's motions. Unlike $\calD\UD{A}{B}$ it is also independent of the parametrization
of $\gamma_0$. In geometrical terms it encodes the information about
how a past light cone with an infinitesimally displaced vertex is positioned with respect to the past light cone originating in $\calO$. It can be obtained independently by solving and 
ODE along $\gamma_0$ with initial conditions at $\calO$. In a parallel propagated Sachs frame they take the form of
\bea
\frac{\dd^2}{\dd \lambda^2}\, m\UD{A}{\mu} - R\UD{A}{\alpha\beta B}\,p^\alpha\,p^\beta\,m\UD{B}{\mu} &=&  R\UD{A}{\alpha\beta\mu}\,p^\alpha\,p^\beta \\
m\UD{A}{\mu}(\lambda_\calO) &=& 0 \\
\frac{\dd}{\dd\lambda}\, m\UD{A}{\mu}(\lambda_\calO) &=& 0.
\eea
This way we have completely split the dependence of $\delta_\calO r^A$ in (\ref{eq:posdrift}) into the dependence on the observer's and emitter's motion (via $u_\calO^\mu$, $u_\calE^\mu$, $z$ and $w_\calO^\mu$)
and on the spacetime geometry (via $\calD\UD{A}{B}(\lambda_\calE)$, $m\UD{A}{\mu}(\lambda_\calE)$ and the parallel transport).

\subsection{Redshift drift}

We will now relate the position drift to the redshift drift. A relation of this kind was first pointed out in \cite{hasse-perlick} in the context of parallax-free 
cosmological models: the authors proved that models in which observers see no change of relative positions of the distant objects on the sky must also posses a redshift potential 
function. In this section we will derive a general formula expressing 
the redshift drifts in terms of the observer's and emitter's motion, the Riemann tensor along the null geodesic and the position drift derived above.

We begin by noting that
\bea
\ln(1+z) &=& \ln\left(p_\mu\,u_\calE^\mu\right) - \ln\left(p_\mu\,u_\calO^\mu\right). \label{eq:ln1plusznotused}
\eea
It is natural to evaluate the first term at $\calE$ and the second at $\calO$, but if we do the parallel transport of $u_\calE^\mu$ we can also evaluate the second one at $\calO$:
\bea
\ln(1+z) &=& \ln\left(p_\mu\,\hat u_\calE^\mu \Big|_\calO\right) - \ln\left(p_\mu\, u_\calO^\mu\right) \label{eq:ln1plusz}.
\eea
Note that in our notation we do not specify explicitly where the unhatted vectors like $u_\calO^\mu$ or $u_\calE^\mu$ need to be evaluated, since they are defined in
$\cal O$ and $\cal E$ respectively. In a similar manner we do not specify where $p_\mu$ should be evaluated as this should be obvious from the context. 
Moreover, if we use a parallel-propagated frame along $\gamma$, we do not need to bother with the evaluation point at all, since $p_\mu$ and hatted (parallel-propagated)
vectors are constant in that case.

The reason why we want to use (\ref{eq:ln1plusz}) rather than more natural (\ref{eq:ln1plusznotused}) is that, unlike (\ref{eq:ln1plusznotused}),  differentiating (\ref{eq:ln1plusz})  leads to a formula involving quantities defined only at $\cal O$, like the position drift.
We now apply $\nabla_X$ to the (\ref{eq:ln1plusz}). The second term on the right hand side is straightforward:
\bea
\nabla_X\ln\left(p_\mu\, u_\calO^\mu\right) &=& \frac{1}{p_\mu\,u_\calO^\mu}\left(\nabla_X p_\mu \Big|_\calO\, u_\calO^\mu + p_\mu\,\nabla_X u_\calO^\mu\right).  \nonumber
\eea
The second term can be expressed using the observer's 4-acceleration, just like in the previous section, while the first one can be re-expressed using (\ref{eq:commutation}):
\bea
\nabla_X\ln\left(p_\mu\,u_\calO^\mu\right) &=& \frac{1}{p_\mu\,u_\calO^\mu}\left(\nabla_p X_\mu \Big|_\calO\, u_\calO^\mu + p_\mu\,w_\calO^\mu\right).  \nonumber
\eea
Finally, the term involving $\nabla_p X_\mu$ can be related to the position drift using the results of the previous section.

The derivative of the first term in (\ref{eq:ln1plusz}) requires a bit more care. Obviously, we may repeat the first step and obtain
\bea
\nabla_X\ln\left(p_\mu\, \hat u_\calE^\mu\right) &=& \frac{1}{p_\mu\,u_\calE^\mu}\left(\nabla_X p_\mu \Big|_\calO\, \hat u_\calE^\mu\Big|_\calO
+ p_\mu\,\nabla_X \hat u_\calE^\mu\Big|_\calO\right).  \nonumber
\eea
Again, we may substitute $\nabla_p X_\mu$ for $\nabla_X p_\mu$ in the first term, but the second one is not the emitter's 4-acceleration. It may nevertheless be related to
$w_\calE^\mu$ by the following reasoning: at $\calE$ we have obviously
\bea
p_\mu\,\nabla_X u_\calE^\mu = \frac{1}{1+z}\, p_\mu\,\nabla_{u_\calE} u_\calE^\mu = \frac{1}{1+z}\,p_\mu\,w_\calE^\mu. \label{eq:IDpDXu}
\eea 
On the other hand, we may calculate the derivative of this product along the fiducial null geodesic:
\bea
\nabla_p\left( p_\mu\,\nabla_X \hat u_\calE^\mu\right) = p_\mu\,\nabla_p\left( \nabla_X \hat u_\calE^\mu\right) =
p_\mu\,\nabla_X\left( \nabla_p \hat u_\calE^\mu\right) + p_\mu\,R\UD{\mu}{\nu\alpha\beta}\,\hat u_\calE^\nu\, p^\alpha\,X^\beta.  \nonumber
\eea
The second equality follows from the commutation of $X^\mu$ and $p^\mu$ and the definition of the Riemann tensor. 
The first term in the last expression vanishes because $\hat u_\calE^\mu$ is parallel-propagated, so we are left with 
\bea
\nabla_p\left( p_\mu\,\nabla_X \hat u_\calE^\mu\right) =   R_{\mu\nu\alpha\beta}\,p^\mu\,\hat u_\calE^\nu\,p^\alpha\,X^\beta. \label{eq:ODEpDXu}
\eea
(\ref{eq:IDpDXu}) and (\ref{eq:ODEpDXu}) constitute the initial data and an ODE respectively. It can be solved to yield
\bea
p_\mu\,\nabla_X \hat u_\calE^\mu\Big|_\calO =  \frac{1}{1+z}\,p_\mu\,w_\calE^\mu + \int_{\lambda_\calE}^{\lambda_\calO} 
R_{\mu\nu\alpha\beta}\,p^\mu\,\hat u_\calE^\nu\,p^\alpha\,X^\beta \,\dd\lambda. \nonumber
\eea 
Note that the first term can also be evaluated at $\calO$ if we use the parallel-propagated $\hat w_\calE^\mu$.
We may now substitute these results to (\ref{eq:ln1plusz}), rearrange the terms and obtain
\bea
\nabla_X \ln \left(1+z\right) &=& \frac{1}{p_\sigma \,u_\mathcal{O}^\sigma}\left.\left(\left(\frac{1}{(1+z)^2} \hat w_\mathcal{E}^\mu - 
w_\mathcal{O}^\mu\right)\,\right|_{\mathcal{O}} p_\mu +
\frac{1}{1+z}\int_{\lambda_\mathcal{E}}^{\lambda_\mathcal{O}} R_{\alpha\beta\mu\nu}\,p^\alpha\,\hat u_\mathcal{E}^\beta\,p^\mu\,X^\nu \dd \lambda \right. \nonumber\\
&&\left.+\nabla_p X^\mu(\lambda_\calO)\,\left.\left(\frac{1}{1+z} \hat u_\mathcal{E} - u_\mathcal{O}\right)_\mu\,\right|_{\mathcal{O}}\right).  \nonumber
\eea
In the last term both vectors are perpendicular to $p^\mu$, so the product can be evaluated on the $\calP$ space: 
\bea
\nabla_X \ln \left(1+z\right) &=& \frac{1}{p_\sigma \,u_\mathcal{O}^\sigma}\left.\left(\left(\frac{1}{(1+z)^2} 
\hat w_\mathcal{E}^\mu - w_\mathcal{O}^\mu\right)\,\right|_{\mathcal{O}} p_\mu +
\frac{1}{1+z}\int_{\lambda_\mathcal{E}}^{\lambda_\mathcal{O}} R_{\alpha\beta\mu\nu}\,p^\alpha\,\hat u_\mathcal{E}^\beta\,
p^\mu\,X^\nu \dd \lambda \right. \nonumber\\
&&\left.+\nabla_p X^A(\lambda_\calO)\,\left.\left(\frac{1}{1+z} \hat u_\mathcal{E} - u_\mathcal{O}\right)_A\,\right|_{\mathcal{O}}\right).
\label{eq:zdrift}
\eea
The resulting formula expresses the total redshift drift by the line-of-sight component of the 4-acceleration difference, evaluated at the observation point, an integral of a component of the Riemann tensor along the geodesic
and a term involving the product of $\nabla_p X^A(\lambda_\calO)$
 and the perpendicular component of the 4-velocity difference.  Note that the Riemann tensor term is insensitive to adding $C\,p^\mu$ to $X^\mu$, therefore there is no need
 to know the $p^\mu$ component of the observer's time vector and it
 is sufficient to use formulas  (\ref{eq:mGDEP}-\ref{eq:mID2P}) and  (\ref{eq:phiwhole}-\ref{eq:totalX}) from Section \ref{sec:Xvect} to evaluate it.
 % CHANGED SINCE SUBMISSION %
 
 Using (\ref{eq:dphiA}) the last term can be re-expressed as a quadratic function involving the perpendicular component 4-velocity difference and $m^A$:
 \bea
  \nabla_p X^A(\lambda_\calO)\,\left.\left(\frac{1}{1+z} \hat u_\mathcal{E} - u_\mathcal{O}\right)_A\,\right|_{\mathcal{O}} &=&
   \left.\left(\frac{1}{1+z} \hat u_\mathcal{E} - u_\mathcal{O}\right)_A\,\right|_{\mathcal{O}}\,{\calD^{-1}}(\lambda_\calE)\UD{A}{B}\nonumber\\
   &&\cdot\left(
   \left.\left(\frac{1}{1+z}  u_\mathcal{E} - \hat u_\mathcal{O}\right)^B - m^B\right)\right|_\mathcal{E} \label{eq:quadrat1}.
 \eea
 Note that ${\calD^{-1}}(\lambda_\calE)\UD{A}{B}$ is a non-local operator in which the first index is contracted with a vector from $\calP_\mathcal{O}$ while 
 the second one with a vector from $\calP_\mathcal{E}$, hence two different evaluation points for the 4-velocity difference.
 (\ref{eq:zdrift}) and (\ref{eq:quadrat1}) yield the general redshift drift in terms of the 
 4-velocities, 4-accelerations, the Jacobi matrix and the Riemann tensor along the null geodesic.

 Alternatively we may express the last term via the position drift using (\ref{eq:posdrift}):
 \bea
  \nabla_p X^A(\lambda_\calO)\,\left.\left(\frac{1}{1+z} \hat u_\mathcal{E} - u_\mathcal{O}\right)_A\,\right|_{\mathcal{O}} &=&
   \left(p_\sigma\,u_\calO^\sigma \right)^2\,\left.\left(\widehat{\delta_\calO r^A} - \hat w_\calO^A\right)\right|_\calE \nonumber \\
     && \cdot\left(\calD(\lambda_\calE)_{AB}\left.\left(\doo r^B - w_\calO^B\right)\right|_\mathcal{O} + m_A \right), \label{eq:quadrat2}
 \eea
 which together with (\ref{eq:zdrift}) gives a (non-perturbative) identity involving the values of the redshift and position drifts.

\subsection{Jacobi matrix drift}

We begin the discussion by stating two important  results. 
\begin{theorem} \label{thm:Dparallel}
Let $\gamma_{\tau}$ be a smooth, one-parameter family of curves (not necessary geodesic) obtained by dragging the fiducial curve $\gamma_0$ by the vector field $X^\mu$, just like in Sec. \ref{sec:geometricoptics}. Let $\iota^\mu(\tau, \lambda)$ be a one-parameter family of solutions of the equation of parallel propagation
 along curves $\gamma_\tau$,
depending on $\tau$ in a smooth way, i.e. let $\iota^\mu$ satisfy the for all $\tau$ and $\lambda$
\bea
\nabla_p \iota^\mu &=& 0 \label{eq:thmparallel}\\
\iota^\mu(\lambda_\calO) &=& A^\mu(\tau) \label{eq:thmID}
\eea
with the second equation giving the initial condition for the first one. Then the derivative $\nabla_X \iota^\mu$ satisfies the inhomogeneous parallel propagation equation with initial data of the form
\bea
\nabla_p \left(\nabla_X \iota^\mu\right) &=& -R\UD{\mu}{\nu\alpha\beta}\,\iota^\nu\,X^\alpha\,p^\beta \label{eq:thmDparallel} \\
\nabla_X \iota^\mu(\lambda_\calO) &=& \delta_\calO A^\mu - A^\nu\,u_{\calO\,\nu}\,w_\calO^\mu + A^\nu\,w_{\calO\,\nu}\,u_\calO^\mu. \label{eq:thmDID}
\eea
\end{theorem}
The proof is fairly straightforward: we differentiate (\ref{eq:thmparallel}--\ref{eq:thmID}) using $\nabla_X$. In the first equation we then 
change the order of covariant derivatives, applying the definition of the
Riemann tensor using commutation of $\nabla_p$ and $\nabla_X$ and taking into account that $[p,X]=0$. In the second equation we simply apply the definition of
$\delta_\calO$ to $\nabla_X \iota^\mu$.

The second theorem is a related result concerning the GDE:
\begin{theorem} \label{thm:DGDE}
Let $\gamma_{\tau}$ be a smooth, one-parameter family of geodesics obtained by dragging the fiducial geodesic $\gamma_0$ by the vector field $X^\mu$, just like in Sec. \ref{sec:geometricoptics}. Let $\xi^\mu(\tau, \lambda)$ be a one-parameter family of solutions of the GDE
 along curves $\gamma_\tau$,
depending on $\tau$ in a smooth way, i.e. let $\xi^\mu$ satisfy the for all $\tau$ and $\lambda$
\bea
\calG[ \xi]^\mu &=& 0  \nonumber\\
\xi^\mu(\lambda_\calO) &=& A^\mu(\tau) \label{eq:GDEid1} \\
\nabla_p\xi^\mu(\lambda_\calO) &=& B^\mu(\tau)  \label{eq:GDEid2} ,
\eea
with the last two equation giving the initial values. Then the derivative $\nabla_X \xi^\mu$ satisfies the inhomogeneous GDE
\bea
\calG[\nabla_X\xi]^\mu &=& \calM\UD{\mu}{\nu}\,\xi^\nu + \calN\UD{\mu}{\nu}\,\nabla_p\xi^\nu
\label{eq:VGDEE}
\eea
with the inhomogeneity given by
\bea
\calM\UD{\mu}{\nu} &=& -(\nabla_\alpha R\UD{\mu}{\nu\rho\sigma})\,p^\alpha\,X^\rho\,p^\sigma + (\nabla_\alpha R\UD{\mu}{\beta\rho\nu})\,X^\alpha\,p^\beta\,p^\rho + 
2R\UD{\mu}{\beta\sigma\nu}\,(\nabla_p X^\beta)\,p^\sigma \label{eq:Mdef}\\
\calN\UD{\mu}{\nu} &=& -2R\UD{\mu}{\nu\rho\sigma}\,X^\rho\,p^\sigma. \label{eq:Ndef}
\eea
Additionally, the initial data for $\nabla_X \xi^\mu$ reads
\bea
\nabla_X \xi^\mu(\lambda_\calO) &=& \delta_\calO A^\mu - A^\nu\,u_{\calO\,\nu}\,w_\calO^\mu + A^\nu\,w_{\calO\,\nu}\,u_\calO^\mu \label{eq:DGDEid1}\\
\nabla_p(\nabla_X  \xi^\mu) &=& \delta_\calO B^\mu - B^\nu\,u_{\calO\,\nu}\,w_\calO^\mu + B^\nu\,w_{\calO\,\nu}\,u_\calO^\mu - R\UD{\mu}{\nu\alpha\beta}\,\xi^\nu\,X^\alpha\,p^\beta.\label{eq:DGDEid2}
\eea 
\end{theorem}
The proof is conceptually similar to the proof of the previous theorem, but significantly more involved computationally. A sketch of the derivation is given in the Appendix 
\ref{appendix}. Note that equations (\ref{eq:VGDEE}--\ref{eq:Ndef}) involve the $X^p$ component of the $X^\mu$ vector field, unlike the equations for the redshift and position drifts.

The theorems formulate ODE systems we may use to obtain the first order approximation of the solution of the parallel propagation equation or GDE at time $\tau = \tau_0 + \dd \tau$
given the solution at $\tau = \tau_0$.
With these two statements in hand, we may now proceed to the problem of the Jacobi matrix drift. Consider the observer's adapted frame 
$\left(u_\calO^\mu,e\DU{A}{\mu},r^\mu\right)$ parallel propagated along $\gamma_0$ yielding $\left(\hat u_\calO^\mu,\hat e\DU{A}{\mu},\hat r^\mu\right)$.
Let $(\hat \omega\UD{A}{\mu})$ be the parallel propagated Sachs co-frame, i.e. a pair of vectors satisfying
\bea
 \hat \omega\UD{A}{\mu}\,\hat e\DU{B}{\mu} &=& \delta\UD{A}{B}  \nonumber\\
 \hat \omega\UD{A}{\mu}\,p^\mu &=& 0  \nonumber\\
 \hat \omega\UD{A}{\mu}\,\hat u_\calO^\mu &=& 0.  \nonumber
\eea
The components of the Jacobi matrix in this frame can be now expressed as
\bea
 \calD\UD{A}{B}(\lambda) = \hat\omega\UD{A}{\mu}\,\eta\DU{B}{\mu},\label{eq:newJacobi}
\eea
where $\eta\DU{1}{\mu}$ and $\eta\DU{2}{\mu}$ are two solutions of the GDE with initial data
\bea
 \eta\DU{B}{\mu}(\lambda_\calO) &=& 0 \nonumber \\
 \nabla_p \eta\DU{B}{\mu}(\lambda_\calO) &=& e\DU{B}{\mu}. \nonumber
\eea
We differentiate (\ref{eq:newJacobi}) using $\nabla_X$:
\bea
 \nabla_X\calD\UD{A}{B} = (\nabla_X\hat\omega\UD{A}{\mu})\,\eta\DU{B}{\mu} + \hat\omega\UD{A}{\mu}\,\nabla_X\eta\DU{B}{\mu}.\label{eq:newJacobi2}
\eea
By the virtue of Theorem \ref{thm:DGDE} the second term can be obtained by solving the inhomogeneous GDE of the form
\bea
 \calG[\nabla_X \eta_A]^\mu &=& \calM\UD{\mu}{\nu}\,\eta\DU{A}{\nu} + \calN\UD{\mu}{\nu}\,\nabla_p\eta\DU{A}{\nu}  \nonumber\\
 \nabla_X \eta\DU{A}{\mu}(\lambda_\calO) &=& 0  \nonumber\\
 \nabla_X (\nabla_p \eta\DU{A}{\mu}) (\lambda_\calO) &=& \delta_\calO e\DU{A}{\mu}  + w_{\calO\,A}\,u_\calO^\mu. \label{eq:DJacobiID2}
\eea
 The Fermi-Walker derivative in the first
term can be calculated if we note that $e\DU{A}{\mu}$ as the two perpendicular vectors of the observer's frame need to be orthogonal to both $u_\calO^\mu$ and $r^\mu$. 
From this it is straightforward to prove that 
\bea
 \delta_\calO e\DU{A}{\mu} = \Omega\UD{B}{A}\,e\DU{B}{\mu} - r^\mu\,\delta_\calO r_A,  \nonumber
\eea
where $\Omega_{AB}$ is an antisymmetric two-by-two matrix generating the rotation of the screen vectors around the direction of observation.
It is pure gauge and can be set to 0 for simplification. In this case (\ref{eq:DJacobiID2}) takes the form of
\bea
 \nabla_X (\nabla_p \eta\DU{A}{\mu}) (\lambda_\calO) &=& - r^\mu\,\delta_\calO r_A + w_{\calO\,A}\,u_\calO^\mu.  \nonumber
\eea

Since $\hat \omega\UD{A}{\mu}$ is parallel-propagated, the first term of (\ref{eq:newJacobi2}) requires the use of Theorem \ref{thm:Dparallel}. Stricly speaking, we 
need to consider the two parallel propagated vectors with raised indices $\hat \omega^{A\,\mu}$, but note that parallel propagation and index raising commute. 
After simple algebraic manipulations we see that $\nabla_X \omega\UD{A}{\mu}$ needs to satisfy 
\bea
\nabla_p \left(\nabla_X \hat\omega\UD{A}{\mu}\right) &=& R\UD{A}{\mu\alpha\beta}\,X^\alpha\,p^\beta \label{eq:omegaa} \\
\nabla_X \omega\UD{A}{\mu}(\lambda_\calO) &=& \delta_\calO \omega\UD{A}{\mu}+ w_{\calO}^A\,u_{\calO\,\mu}. \label{eq:omegaaa}
\eea
Using a similar reasoning as in the case of $e\DU{A}{\mu}$, we show that
we can substitute $\delta_\calO \omega\UD{A}{\mu}$ by $-r_\mu\,\delta_\calO r^A$.
This way we obtain ODE's with initial data at $\calO$ which we need to solve up to $\calE$ and then substitute to (\ref{eq:newJacobi2}).

It turns out, however, that for many purposes we do not need to solve (\ref{eq:omegaa}-\ref{eq:omegaaa}) explicitly. 
Consider the components of $\nabla_X \hat\omega\UD{A}{\mu}$ in the observer's null co-frame $\left(\hat u_{\calO\,\mu},\hat\omega\UD{A}{\mu},p_\mu\right)$:
\bea
 \nabla_X \hat\omega\UD{A}{\mu} = \Psi\UD{A}{B}\,\hat\omega\UD{B}{\mu} + C^A\,\hat u_{\calO\,\mu} + D^A\,p^\mu.  \nonumber
\eea
From the conditions $\hat\omega\UD{A}{\mu}\,\hat\omega^{B\,\mu} = \delta^{AB}$ and $\hat\omega\UD{A}{\mu}\,p^\mu = 0$ it is easy to show
that $\Psi_{AB} = -\Psi_{BA}$ and $C^A = -\frac{\nabla_p X^A}{p_\sigma\,u_\calO^\sigma}$. Moreover, vectors $\eta\DU{A}{\mu}$ are orthogonal to $p^\mu$, so (\ref{eq:newJacobi2})
simplifies to
\bea
 \nabla_X \calD\UD{A}{B} = \hat\omega\UD{A}{\mu}\,\nabla_X\eta\DU{B}{\mu} - (\nabla_p X^A)\,\eta\DU{B}{p} + \Psi\UD{A}{C}\,\calD\UD{C}{B}, \label{eq:Jacobidrift}
\eea
where $\eta\DU{B}{p}$ is the $p^\mu$ component in the decomposition in the null frame and all quantities are evaluated at $\calE$. Note that
 the last term generates a rotation of the Jacobi matrix around the line of sight, keeping the shapes of the lensed objects fixed. 
 Unlike the second term, involving $\nabla_p X^A$, it cannot be expressed by the data we already know from solving the previous ODE's -- one 
 really needs to solve (\ref{eq:omegaa}--\ref{eq:omegaaa}) 
 in order to obtain the matrix $\Psi_{AB}$. 
 We can neglect it however as long as we are only interested in the drift of the size and the distortion of the images, not in the change of their orientation. The only other relevant term
 of $\nabla_X \hat\omega\UD{A}{\mu}$, i.e. $C^A$, turns out to be expressible by $\nabla_p X^A$, which we already know. Therefore there is no need to solve (\ref{eq:omegaa}--\ref{eq:omegaaa}).

 \subsection{Area and luminosity distance drift}
 
Finally, we calculate the derivative of the area and luminosity distances given by (\ref{eq:Dangdef}) and (\ref{eq:etherington}). Differentiating the log of $D_{ang}$ we obtain
\bea
\nabla_X \ln D_{ang} &=& \frac{1}{2}\,\left(\nabla_X\calD\UD{A}{B}\cdot{\calD^{-1}}\UD{B}{A}\right)+\frac{1}{p_\mu\,u_\calO^\mu}
\left(\nabla_p X_\sigma(\lambda_\calO)\,u_\calO^\sigma + p_\sigma\,w_\calO^\sigma\right). \label{eq:Dangdrift}
\eea
We now can substitute (\ref{eq:Jacobidrift}) to the first term. We notice that the rotation generating term with $\Psi_{AB}$  drops out, so
\bea
\nabla_X \ln D_{ang} &=& \frac{1}{2}\,\left.\left(\left(\hat\omega\UD{A}{\mu}\,\nabla_X\eta\DU{B}{\mu} - (\nabla_p X^A)\,\eta\DU{B}{p}\right)\right|_\calE \cdot{\calD^{-1}}(\lambda_\calE)\UD{B}{A}\right)\nonumber\\
&&+\frac{1}{p_\mu\,u_\calO^\mu}
\left(\left(\nabla_p X_\sigma(\lambda_\calO)\right)\,u_\calO^\sigma + p_\sigma\,w_\calO^\sigma\right)\label{eq:Dangdrift2}
\eea
and we can again skip solving (\ref{eq:omegaa}--\ref{eq:omegaaa}).
 
 Differentiating the logarithm of the Etherington's reciprocity relation (\ref{eq:etherington}) we obtain an expression for the luminosity distance drift
\bea
\nabla_X \ln D_{lum} = \nabla_X \ln D_{ang} + 2\nabla_X \ln (1+z), \label{eq:Dlumdrift}
\eea
into which we need to insert (\ref{eq:zdrift}).

 \section{Discussion}
 
 \subsection{Dependence on the observer, emitter and spacetime}
 
 The null geodesic between $\calO$ and $\calE$, if it exists at all, does not depend on the 4-velocities of the emitter and the observer. This trivial observation has 
 interesting consequences. It follows that the covariant derivative 
 $\nabla_p X^A$ can only depend on the time derivative of the positions of both the emitter and the observer, i.e. their 4-velocities, and the geometry of the spacetime
 between them, represented by the values of the Riemann tensor along $\gamma_0$, see (\ref{eq:dphiA}):
 \bea
  \nabla_p X^A \equiv \nabla_p X^A (R\UD{\mu}{\nu\alpha\beta},u_\calO^\mu,u_\calE^\mu).  \nonumber
 \eea
 Note that the dependence on $u_\calE^\mu$ is only via its the transversal components, the line-of-sight component does not matter.
 On the other hand, the closely related Fermi-Walker derivative of the position on the sky depends also on the 4-acceleration of the observer because it
 involves the aberration effects, 
 see (\ref{eq:posdrift}):
 \bea
 \doo r^A \equiv \doo r^A  (R\UD{\mu}{\nu\alpha\beta},u_\calO^\mu,u_\calE^\mu,w_\calO^\mu).  \nonumber
 \eea
 The redshift drift obviously depends on both 4-velocities, but also on the line-of-sight 4-accelerations of both observer and the emitter, as can be seen in equation (\ref{eq:zdrift}).
 \bea
 \nabla_X \ln(1+z) \equiv \nabla_X \ln(1+z)  (R\UD{\mu}{\nu\alpha\beta},u_\calO^\mu,u_\calE^\mu,w_\calO^\mu,w_\calE^\mu).  \nonumber
 \eea
 As we noted in Section \ref{sec:dependence1}, the angular distance does not depend on the 4-velocity of the emitter. Consequently, its time derivative, 
 expressed in (\ref{eq:Dangdrift}-\ref{eq:Dangdrift2}), depends on 
 the Riemann tensor, its derivative, both 4-velocities, but only on the 4-acceleration of the observer:
 \bea
 \nabla_X D_{ang} \equiv \nabla_X D_{ang}  (\nabla_\gamma R\UD{\mu}{\nu\alpha\beta},R\UD{\mu}{\nu\alpha\beta},u_\calO^\mu,u_\calE^\mu,w_\calO^\mu).  \nonumber
 \eea
On the other hand, the luminosity distance  depends on all the quantities involved, as seen in equation (\ref{eq:Dlumdrift})
 \bea
 \nabla_X D_{lum} \equiv \nabla_X D_{lum}  (\nabla_\gamma R\UD{\mu}{\nu\alpha\beta},R\UD{\mu}{\nu\alpha\beta},u_\calO^\mu,u_\calE^\mu,w_\calO^\mu,w_\calE^\mu).  \nonumber
 \eea
 
 \subsection{Relation to the classical cosmological parallax results}
 
 We would like to relate the position drift formulas (\ref{eq:dphiA}, \ref{eq:posdrift}) to the classical results  of McCrea \cite{mccrea}, 
  Kasai \cite{kasai}, Rosquist \cite{rosquist}, Perlick \cite{perlick} and the more recent ones by R\"as\"anen \cite{rasanen} or Hellaby and Walters \cite{Hellaby:2017soj}.

  In \cite{kasai, rasanen} the position drift is expressed as a function of the gradient of the vector $k^\mu$ tangent to a congruence of null geodesics focused on the 
  \emph{emitter's} worldline. The dependence on the observer's motion is thus implicitly present in the tangent vector gradient.   
  We therefore begin by relating the right hand side of equation (\ref{eq:dphiA}) to $\nabla_\mu k^\nu$. Consider (\ref{eq:dphiA}) for a fixed $\calE$ and $u_\calE^\mu$, 
  describing the emitter, but variable $\calO$ and $u_\calO^\mu$.  By construction, $X^\mu(\lambda)$ describes in this case an infinitely thin slice of a 
  congruence of null geodesics originating on the emitter's worldline. The tangent vector $p^\mu$ is equal to  $k^\mu$ defined above. At $\calO$ we thus have 
  \bea
  \nabla_p X^\mu(\lambda_\calO)  = \nabla_X p^\mu(\lambda_\calO)  = X^\nu\,\nabla_\nu k^\mu = u_\calO^\nu \,\nabla_\nu k^\mu. \label{eq:dX1}
\eea   
  Now we rewrite the right hand side of (\ref{eq:dphiA}). Recall that $m^\mu$ from (\ref{eq:mGDEP}--\ref{eq:mID2P}) is a linear function of the observer's 4-velocity, i.e. 
  $m^A \equiv m\UD{A}{\mu}\,u_\calO^\mu$, where $m\UD{A}{\mu}$ is a linear operator from the tangent space at $\calO$ to $\calP$ at $\calE$.
  Thus,
  \bea
   \nabla_p X^A(\lambda_\calO) = -{\mathcal{D}^{-1}}\UD{A}{B}\left(\lambda_\mathcal{E}\right)\,\left(\Pi\UD{B}{\mu} + m\UD{B}{\mu} \right) u_\calO^\mu, \label{eq:dX2}
  \eea
  where $\Pi\UD{B}{\mu}$ denotes the pullback to $\calP$ of the projection operator $\Pi\UD{\mu}{\nu}$ 
  \bea
  \Pi\UD{\mu}{\nu} = \delta\UD{\mu}{\nu} - \frac{1}{p_\sigma\,u^\sigma_\calE}\,u_\calE^\mu\,p_\nu  \label{eq:Piop}
  \eea
 projecting to the subspace $p^\perp$ along $u_\calE^\mu$.  Finally, noting that (\ref{eq:dX1}) and (\ref{eq:dX2}) must both hold for any future-pointing, 
 normalized timelike vector $u_\calO^\mu$, we obtain
 \bea
  \nabla_\nu k^A = -{\mathcal{D}^{-1}}\UD{A}{B}\left(\lambda_\mathcal{E}\right)\,\left(\Pi\UD{B}{\nu} + m\UD{B}{\nu} \right),  \nonumber
 \eea
 with $\Pi\UD{\mu}{\nu}$ depending on the emitter's motion. 
If we restrict the lower index to $\calP$, the expression above simplifies to
\bea
  \nabla_C k^A = -{\mathcal{D}^{-1}}\UD{A}{B}\left(\lambda_\mathcal{E}\right)\,\left(\delta\UD{B}{C} + m\UD{B}{C} \right), \nonumber
 \eea 
 losing this way any dependence on the emitter's 4-velocity. 
 This expression can be further decomposed to the expansion and shear of the null congruence in the standard manner, $\nabla_A k_B = \frac{1}{2}\,\theta q_{AB} + \sigma_{AB}$.
 We have thus shown that (\ref{eq:dphiA}) provides a general expression for the perpendicular components of the gradient of the null tangent vector of the congruence.
 
In order to make a more direct connection to the classical results, we need to consider a situation when the observer, on top of his large scale, ``cosmological'' motion 
  given by $U_\calO^\mu$, 
  has a time-dependent peculiar motion with characteristic time and distance scales much shorter than  the characteristic scales connected with the beam of light coming from
  the emitter and the curvature scale of the background metric.
  
 Introduce a locally flat coordinate system $(y^\mu)$ around $\calO$, in which $y^\mu = 0$ corresponds to $\calO$, light signals propagate along $y^3$ and such that
 \bea
  g_{\mu\nu}(y^\alpha) = \eta_{\mu\nu} + O\left(|y|^2\right). \nonumber
 \eea
 All subsequent equations will be expressed in this particular coordinate system. Note that the first order Taylor series for $k^\mu$ has the form of
 \bea
  k^\mu = k^\mu(\calO) + \nabla_\nu k^\mu(\calO)\,y^\nu +  O\left(|y|^2\right). \label{eq:taylor}
 \eea
 Assume now that the observer's motion can be described as a fast, non-relativistic perturbation on top of 
 the large-scale, secular motion described by $U_\calO^\mu$, i.e.
 $y^\mu(\tau) = \tau \,U_\calO^\mu + z^\mu(\tau)$, where $U_\calO^\mu$ is considered constant and $z^\mu$ is perpendicular to $U_\calO^\mu$. 
 In this case,
 \bea
  u_\calO^\mu(\tau) = U_\calO^\mu + \dot z^\mu(\tau)  \nonumber
 \eea
with $\dot{z}_\mu\,U_\calO^\mu = 0$.  
  
 Decompose now the observer's peculiar motion in the  null frame $(U_\calO^\mu, e\DU{A}{\mu}, p^\mu)$, associated with observer $U_\calO^\mu$:
 \bea
  z^\mu = z^A\,e\DU{A}{\mu} + z^r \,\left( U_\calO^\mu + \frac{1}{p_\sigma\,U_p^\sigma}\,p^\mu \right).  \nonumber
 \eea
 We consider the change of direction of the incoming photons (disregarding or subtracting the aberration effects, as in \cite{kasai, rosquist, perlick}) in comparison with
 the direction registered at $\calO$. From (\ref{eq:taylor}) we obtain
 \bea
  \Delta r^A = r^A(z^\mu) - r^A(\calO) \approx \frac{1}{p_\sigma\,U_\calO^\sigma} \left(z^C\,\nabla_C k^A + (\tau + z^r)\, U_\calO^\mu \nabla_\mu k^A \right). \label{eq:parallaxnew}
 \eea
  The apparent position change is a sum of two effects. The first one is solely due to the observer's peculiar motion and is proportional to the perpendicular
  displacement $z^A$  of the observer. It is independent of the emitter's 4-velocity, although in case of strongly asymmetric lensing it may depend on the direction of
  the baseline. The second term contains the effects of both observer's and the emitter's motion as well as time- and position-dependent light bending:
  \bea
   U_\calO^\mu \nabla_\mu k^A  =   \left(\frac{1}{1+z}\,u_\mathcal{E} - \hat U_\mathcal{O} \right)^A - m\UD{A}{\mu}\,U_\calO^\mu.  \nonumber
  \eea
  
  Recall that the first term in (\ref{eq:parallaxnew}) is the basis of the classical definition of the parallax distance \cite{mccrea, kasai, rosquist, rasanen}:
  \bea
   D_{par} = \frac{2 p_\sigma\,u_\calO^\sigma}{\nabla_A k^A}, \nonumber
  \eea 
  which uses the trace of $\nabla_A k^B$, i.e. the expansion, to define a baseline direction-independent quantity. Just like $D_{ang}$, $D_{par}$ depends on the observer's 4-velocity 
  via the product $p_\sigma\,u_\calO^\sigma$, but not on $u_\calE^\mu$, i.e. $D_{par} \equiv D_{par}\left(R\UD{\mu}{\nu\alpha\beta},u_\calO^\mu\right)$. 
  The main problem with this definition of cosmological distance is the difficulty of a measurement.
  
  Firstly, the measurement must be done along at least two different baselines in order to determine the effect along two orthogonal directions and 
  obtain the value of the $\nabla_A k^A$.
  Secondly, one should in principle conduct the measurements at the ends of each baseline simultaneously (as noted in the $U_\calO^\mu$ frame) 
  in order to get rid of the second term in (\ref{eq:parallaxnew}). Indeed, for a baseline lying on the perpendicular plane through $\calO$ we have $z^r=0$ and 
  for  a  measurement simultaneous with $\calO$ we have $\tau = 0$, so the apparent position difference $\Delta r^A$ depends only on the position along the baseline $z^A$. 
  But this means that we need to compare measurements at two points of spacelike separation, which is very difficult in the cosmological setting.
  Obviously, missions like Gaia can only make measurements along a timelike worldline and, as was noted in \cite{rasanen},
  the observed parallax may be different because the second term will contribute as well.

 If we assume a large time scale difference between the two effects, 
  we may approximate the second term as linear in $\tau$ and try to separate it out from the total signal. 
  On the other hand, in  an FLRW model in which both emitter and observer follow the Hubble flow, this secular term vanishes identically and the 
 total effect is just due to the first term. The net result in this case agrees with \cite{kasai}.
 
 In \cite{Hellaby:2017soj} the authors take a complementary approach and consider the null congruence focused at the 
 \emph{observer's} worldline. Consequently the position drift formula they arrive at is composed of a linear expression in $u_\calE^\mu$ multiplied by $(1+z)^{-1}$. 
 It is straightforward to recast (\ref{eq:dphiA}) in a similar way by algebraic manipulations: we need to define the projection 
 operator $\Sigma\UD{\mu}{\nu}$ analogous to (\ref{eq:Piop}), but with
 $u^\mu_\calO$ playing the role of $u^\mu_\calE$:
 \bea
 \Sigma\UD{\mu}{\nu} = \delta\UD{\mu}{\nu} - \frac{1}{p_\sigma\,u^\sigma_\calO}\,u_\calO^\mu\,p_\nu. \nonumber
 \eea
 Then it is easy to prove that
 \bea
\nabla_p X^A = \frac{1}{1+z}\, {\calD(\lambda_\calE)^{-1}}\UD{A}{B} \left( \Sigma\UD{B}{\nu} - m^B\,p_\nu\right)\,u_\calE^\nu.
 \eea 
 Making a direct comparison would require taking into account the aberration effects and comparing the resulting expression with the 
 GDE-transported basis from \cite{Hellaby:2017soj}; we skip it for brevity.
 
 \subsection{Numerical applications}
 The equations derived in this paper form a hierarchy. This structure is very convenient from the point of view of numerical application, because we may use the 
 previous results to obtain the new ones, improving this way the computational cost. 
 The correct order of calculations is the following: given a solution of the Einstein equations and two points $\calE$ and $\calO$ connected by a null geodesics,
 we first solve the parallel propagation equation for a Sachs frame, the solve (\ref{eq:JacMat}) in order to obtain the Jacobi matrix and the angular and luminosity distances. 
 We then solve (\ref{eq:mGDEP}) to obtain $m^A$ and from these data we can calculate the position drift (\ref{eq:posdrift}).  With this data available we only need to evaluate one integral
 of the Riemann tensor to obtain the redshift drift via (\ref{eq:zdrift}). Finally, for the Jacobi matrix drift and the angular and luminosity distance drifts we need  
 the $X^p$ component of $X^\mu$. We get it by solving the $p$ component of (\ref{eq:mGDE}). With this data we can solve equations (\ref{eq:newJacobi}--\ref{eq:omegaaa})
 in order to obtain $\nabla_X \calD\UD{A}{B}$ and the distances drifts.

 \subsection{Summary}
 
 We have presented a new method to obtain the optical drift effects in an arbitrary spacetime. The method is based on the null geodesic deviation equation and can be
 considered an extension of the standard geometric optics formalism in GR to the time derivatives of the optical observables. In particular, we have provided a new
 expression for the position drift $\delta_\calO r^A$ in terms of the 4-velocities of the objects involved and the Jacobi matrix, expressing this way the relation between the 
 gravitational lensing and the position drift. We have also derived an expression for the redshift drift
 as a function of the position drift as well as the 4-accelerations of the observer and the emitter. It can be seen as a non-perturbative identity the two drifts need to 
 satisfy, analogous to the Etherington's duality relation. Finally, we have shown how one can obtain the Jacobi matrix drift and the angular and luminosity distance drifts
 using the first derivative of the GDE. We have also discussed the relation of the formulas presented in this paper to the previous results in the field.

 \section*{Acknowledgements}
 The work was supported by the National Science Centre, Poland (NCN) via the SONATA BIS programme, grant No~2016/22/E/ST9/00578 for the project
\emph{``Local relativistic perturbative framework in hydrodynamics and general relativity and its application to cosmology''}. MK would like to thank 
Filip Ficek, Andrzej Krasi\'nski and Syksy R\"as\"anen for useful discussions and comments.

\appendix 
\section{Proof of Theorem \ref{thm:DGDE}}\label{appendix}

First, we prove equations (\ref{eq:VGDEE}--\ref{eq:Ndef}). We begin by differentiating (\ref{eq:GDEfirst}) with respect to $X$: 
\bea
\nabla_X\left(\nabla_p\nabla_p \xi^\mu - R\UD{\mu}{\alpha\beta\nu}\,p^\alpha\,p^\beta\,\xi^\nu\right) = 0. \label{eq:derivationVGDE0}
\eea
In the first term we would like to change the order of covariant differentiation in order to obtain $\nabla_p\nabla_p\left(\nabla_X \xi^\mu\right)$. 
We can do it in three steps. We first note that
\bea
\nabla_X\left(\nabla_p\nabla_p \xi^\mu\right) = X^\alpha\,p^\beta\,p^\gamma\,\nabla_\alpha\nabla_\beta\nabla_\gamma\xi^\mu + 
2\nabla_p(\nabla_\alpha \xi^\mu)\,\nabla_p X^\alpha + R\UD{\mu}{\sigma\alpha\beta}\,\xi^\sigma\,(\nabla_p X^\alpha)\,p^\beta \label{eq:derivationVGDE1}
\eea
(obtained using the Leibniz rule for the covariant derivative, switching the order of derivatives via the Ricci identity and using $[X,p] = 0$).

Commuting the indices of the third covariant derivative of $\xi^\mu$ with the help of the Ricci identity leads to
\bea
X^\alpha\,p^\beta\,p^\gamma\,\nabla_\alpha\nabla_\beta\nabla_\gamma\xi^\mu  &=& X^\alpha\,p^\beta\,p^\gamma\left(\nabla_\beta\nabla_\gamma\nabla_\alpha\xi^\mu +
(\nabla_\beta R\UD{\mu}{\sigma\alpha\gamma})\,\xi^\sigma \right. \nonumber \\
&&\left. + R\UD{\mu}{\sigma\alpha\gamma}\,\nabla_\beta\xi^\sigma + R\UD{\mu}{\sigma\alpha\beta}\,\nabla_\gamma\xi^\sigma -
R\UD{\sigma}{\gamma\alpha\beta}\,\nabla_\sigma\xi^\mu\right). \label{eq:derivationVGDE2}
\eea
Finally, we prove again by direct differentiation that
\bea
 X^\alpha\,p^\beta\,p^\gamma\,\nabla_\beta\nabla_\gamma\nabla_\alpha\xi^\mu = \nabla_p \nabla_p \left(\nabla_X \xi^\mu\right) - 2\nabla_p\left(\nabla_\sigma\xi^\mu\right)
 \nabla_p X^\sigma - \nabla_p \nabla_p X^\sigma\,\nabla_\sigma \xi^\mu. \label{eq:derivationVGDE3}
\eea
Putting the equations above together we note that the last term of (\ref{eq:derivationVGDE2}) and the last term of (\ref{eq:derivationVGDE3}) cancel out because
by assumptions $X^\mu$ drags geodesics into geodesics, so it needs to satisfy the GDE. Similarly, the second term of (\ref{eq:derivationVGDE1}) and the second term of (\ref{eq:derivationVGDE3}) cancel each other out, therefore
\bea
\nabla_X\left( \nabla_p\nabla_p \xi^\mu \right) &=& \nabla_p\nabla_p\left(\nabla_X \xi^\mu\right) + R\UD{\mu}{\sigma\alpha\beta} \xi^\sigma\,\nabla_p X^\alpha\,p^\beta \\
 &&+ X^\alpha\,p^\beta\,p^\gamma\left((\nabla_\beta R\UD{\mu}{\sigma\alpha\gamma})\,\xi^\sigma + 2R\UD{\mu}{\sigma\alpha\left(\gamma\right.}\,\nabla_{\left.\beta\right)}
 \xi^\sigma\right).  \nonumber
\eea

We have thus obtained the desired expression for the derivative of the first term in brackets in equation (\ref{eq:derivationVGDE0}). It remains now to differentiate the second term
 and collect all the terms. It turns out that we obtain (\ref{eq:VGDEE}) with $\calN\UD{\mu}{\nu}$ given by equation (\ref{eq:Ndef}),
 but $\calM\UD{\mu}{\nu}$ has the form of
 \bea
 \calM\UD{\mu}{\nu} &=& X^\alpha\,p^\beta\,p^\gamma\left(\nabla_\alpha R\UD{\mu}{\beta\gamma\nu} - \nabla_\beta R\UD{\mu}{\nu\alpha\gamma}\right) \nonumber \\
 &&+ (\nabla_p X^\alpha)\,p^\beta \,R\UD{\mu}{\alpha\beta\nu} +   p^\alpha\,(\nabla_p X^\beta) \,R\UD{\mu}{\alpha\beta\nu} - \nabla_p X^\alpha\,p^\beta\,R\UD{\mu}{\nu\alpha\beta}.  \nonumber
 \eea
This can be simplified by applying the cyclic identity for the Riemann tensor 
\bea
R_{\mu\nu\alpha\beta} + R_{\mu\alpha\beta\nu} + R_{\mu\beta\nu\alpha} = 0  \nonumber
\eea
to the last two terms, which leads to (\ref{eq:Mdef}) after small manipulations.

Now we turn to the initial data equations (\ref{eq:DGDEid1}--\ref{eq:DGDEid2}). Differentiating (\ref{eq:GDEid1}) wrt $X^\mu$ we obtain $\nabla_X \xi^\mu = \nabla_X A^\mu$.
We can now apply the definition of the observer's Fermi--Walker derivative (\ref{eq:FWdef}) at the observation point $\calO$ and obtain (\ref{eq:DGDEid1}).

The second initial data equation requires one step more. We begin again by differentiating (\ref{eq:GDEid2}) wrt $X^\mu$ to obtain 
$\nabla_X(\nabla_p \xi^\mu) = \nabla_X B^\mu$. Using the Ricci identity and the commutation of $X^\mu$ and $p^\mu$ we obtain
\bea
\nabla_p\left(\nabla_X \xi^\mu\right) = \nabla_X B^\mu + R\UD{\mu}{\nu\alpha\beta}\,\xi^\nu\,p^\alpha\,X^\beta.  \nonumber
\eea
The final step involves  again applying the definition of the Fermi--Walker derivative to the $\nabla_X B^\mu$ term to obtain (\ref{eq:DGDEid2}).

 \bibliographystyle{jcap/JHEP}
\bibliography{driftpaper}
   \end{document}